\documentclass[smallextended]{svjour3}       
\smartqed  
\usepackage{graphicx}
\usepackage{algorithmic}
\usepackage{algorithm}
\usepackage{verbatim}
\usepackage{subfigure}
\usepackage{hyperref}
\usepackage{tikz}
\usepackage{pgfplots}

\journalname{Computing}

\newcommand{\DEPAS}{\texttt{DEPAS}}

\begin{document}

\pagestyle{empty}
\begin{center}
\vspace{1cm}
\par
{\Large DEPAS: A Decentralized Probabilistic Algorithm  for Auto-Scaling}
\vspace{0.5cm}
\par
Nicol\`o M. Calcavecchia, Bogdan Alexandru Caprarescu, Elisabetta Di Nitto, Daniel J. Dubois, Dana Petcu 

\vspace{3cm}
\par
{\large \textbf{Technical Report n. 2012.5}}

\vspace{0.1cm}
\par
February 10, 2012

\vspace{3cm}
\par
 \includegraphics[scale=0.5]{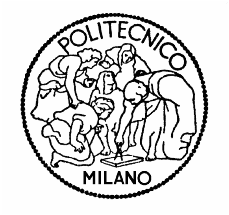}
\vspace{0.5cm}
\par
Dipartimento di Elettronica e Informazione
\vspace{0.1cm}
\par
Politecnico di Milano
\vspace{0.1cm}
\par
Piazza Leonardo da Vinci 32, Milan 20133, Italy

\end{center}
\newpage

\title{\DEPAS{}: A Decentralized Probabilistic Algorithm for Auto-Scaling}
\authorrunning{Calcavecchia, Caprarescu, Di Nitto, Dubois, Petcu}
\titlerunning{\DEPAS{}: A Decentralized Probabilistic Algorithm for Auto-Scaling}

%\thanks{
%}

\author{Nicol\`o M. Calcavecchia \and Bogdan Alexandru Caprarescu \and Elisabetta Di Nitto \and Daniel J. Dubois \and Dana Petcu }

\institute{
Nicol\`o M. Calcavecchia \and Elisabetta Di Nitto \and Daniel J. Dubois \at 
Politecnico di Milano \\ 
Dipartimento di Elettronica e Informazione \\
Piazza Leonardo Da Vinci 32, Milan, Italy  \\
Tel/Fax 00 39 2399 3663  \\
\email{\{calcavecchia, dinitto, dubois\}@elet.polimi.it}
\and
Bogdan Alexandru Caprarescu \and Dana Petcu   \at 
West University of Timisoara \\
IeAT, Faculty of Mathematics and Computer Science \\
Bd. Vasile Parvan 4, RO-300223, Timisoara, Romania\\
Tel/Fax 00 40 256 244834 \\
\email{\{bcaprarescu, petcu\}@info.uvt.ro}
}

\date{}

\maketitle

\begin{abstract}
The dynamic provisioning of virtualized resources offered by cloud computing infrastructures allows applications deployed in a cloud environment to automatically increase and decrease the amount of used resources. This capability is called \emph{auto-scaling} and its main purpose is to automatically adjust the scale of the system that is running the application to satisfy the varying workload with minimum resource utilization. The need for auto-scaling is particularly important during workload peaks, in which applications may need to scale up to extremely large-scale systems.

Both the research community and the main cloud providers have already developed auto-scaling solutions. However, most research solutions are centralized and not suitable for managing large-scale systems, moreover cloud providers' solutions are bound to the limitations of a specific provider in terms of resource prices, availability, reliability, and connectivity.

In this paper we propose \DEPAS{}, a decentralized probabilistic auto-scaling algorithm integrated into a P2P architecture that is cloud provider independent, thus allowing the auto-scaling of services over multiple cloud infrastructures at the same time. Our simulations, which are based on real service traces, show that our approach is capable of: (i) keeping the overall utilization of all the instantiated cloud resources in a target range, (ii) maintaining service response times close to the ones obtained using optimal centralized auto-scaling approaches.
\end{abstract}

\section{Introduction}
\label{sec:introduction}

Cloud computing infrastructures are becoming a valid alternative to buying and maintaining custom on-site IT infrastructures. Some of the reasons are the possibility to pay for the \emph{infrastructural resources} only when they are needed. This flexibility in the amount of resources provided is often called ``elasticity'' and is the basis of the utility computing model~\cite{ec2,sharma2011}. To take full advantage of this flexibility cloud providers give to the customers several tools, called \emph{auto-scaling services}, needed to modify at runtime the amount of used resources. These tools are used to avoid underutilization and overutilization of the cloud resources while still maintaining a good level of quality for the hosted services. This has two advantages: from the point of view of the customer this can be seen as a reduction in the total resources costs, while from the point of view of the cloud provider it allows to serve more customers with the same infrastructure.

In existing cloud infrastructures, the flexibility described above is offered within the individual cloud computing infrastructure, thus leading to the following limitations. First of all the total amount of resources that can be allocated to a single customer is usually limited by the contract and by the total capacity of the cloud provider. Moreover auto-scaling services may have an additional cost. Another problem is that, depending on the current state of the cloud infrastructure (i.e., connectivity) and on the market conditions (i.e., costs) a single cloud provider may not be the most convenient at a certain moment. These limitations are moving current research efforts in cloud computing to the study of efficient ways to exploit more than a single cloud system for the deployment of a service. The capability to \emph{federate} different cloud providers becomes particularly useful when smaller companies want to offer part of their resources for a competitive price in a cloud market. In a scenario like this we would have several cloud computing systems with less capabilities, guarantees, and stability than a single specialized cloud provider.

%%% PART FROM CONTEXT SECTION
The context we are considering in this paper is that of a generic cloud system which offers resources able to host batch services. The characteristics of these batch services is that all the data needed to be processed is given as input, and they return the result when the processing is completed. Compared to a generic service there, in the case of batch service, there is no interaction with the user within the service invocation and the resource that runs the service does not keep the state of the service between invocations. If a state needs to be maintained, then such state is written in a storage utility that is external to the resource (for example a NFS-like distributed storage). Examples of this kind of services are multimedia file conversions (audio and video file transcoding), scientific processing (such as the ones performed on volunteer computing infrastructures, like SETI@Home), strong encryption/decryption, distributed compilation, translation, finding good solutions for NP-hard problems from a model, and any other kind of use in which some input data needs to be transformed in output data in a fast way. There are already companies on the Internet that offer this kind of services to the general public such as~\cite{ankoder,zencoder} for the video file transcoding. These companies have strategic interests in reducing the costs for instantiating their services and in maintaining a good level of satisfaction of the customers that are willing to pay for using them. In the current paper we measure this satisfaction in terms of quality parameters such as \emph{response time} and \emph{drop rate} of the requests (see Section~\ref{sec:results}).
%In this context the functional requirements are not usually a problem since besides licensing issues, every public or privately owned platform is able to offer these services. The real issue is the cost for instantiating these services and the satisfaction of the customers that are willing to pay for using them. 
%%% END OF PART FROM CONTEXT SECTION

The problem we want to solve in this paper is to take advantage of a \emph{cloud federation}  
to avoid the dependence on a single cloud provider, its limitations and costs, while still minimizing the amount of used resources to maintain a good level of quality for the customers.
% in terms of response times and drop rate of their applications.

The solution we propose, denoted with the acronym \DEPAS{} (DEcentralized Probabilistic Algorithm for Auto-Scaling), is \textit{fully decentralized} and \textit{self-organizing}, meaning that none of the cloud providers in the federation have a central role in how the services of the customers are scattered among different providers. Another characteristic of this solution is that it is \textit{self-adapting}, thus reducing the human effort at deployment time in reconfiguring the whole system in case of a serious failure in one cloud provider or in its connectivity. In our approach the self-organizing logic is distributed over all the instantiated cloud resources as a part of the deployed application, and it is autonomously able to make decisions whether to allocate (deallocate) other (existing) resources based on their capacity and on the number of received requests.

We have run our solution in a simulation scenario based on real traces and we have been able to keep the percentage of utilization of the instantiated cloud resources within a given interval (i.e., no resource underutilization, nor resource overutilization) while still obtaining  response times and drop rates comparable to the optimal ones of centralized approaches with full knowledge on the system.

This paper is organized as follows. A state of the art analysis is presented in Section~\ref{sec:state-of-the-art}. 
%Section~\ref{sec:context} presents the context of this work and points out the problem we want to solve. 
Section~\ref{sec:solution} presents our solution, which consists in a system architecture and in a self-organizing algorithm. 
In Section~\ref{sec:results} we show some simulations based on real service demand traces, %Section~\ref{sec:related-work} shows some approaches that are related to the one we are presenting, 
and finally Section~\ref{sec:conclusion} concludes the paper.

\section{State of the art}
\label{sec:state-of-the-art}

In a cloud environment, where resources can be dynamically allocated and deallocated, the scalability of offered services is no longer constrained by the amount of hardware resources that were initially provisioned. Therefore, autonomic resource provisioning in a virtualized environment has emerged as a rich research field situated at the crossroads of autonomic and cloud computing. In this section, we survey several relevant contributions to the field.

A first distinction can be made between provider-centric solutions, whose main objective is to maximize the profit of the provider, and customer-centric solutions which enable the customer to select the best resource configuration that maximizes their needs. In Table \ref{table:related-work-survey}, the commonalities and differences among several contributions, both customer-centric and provider-centric, are surveyed according to three aspects: the autonomic architectural approach (either endogenous or exogenous), the autonomic mechanism used to make adaptation decisions, and the provisioning mechanism. Beyond the criteria used in Table \ref{table:related-work-survey}, we also discuss two emerging directions: SLA-aware provisioning and multi-cloud provisioning. 

The architectural approaches to autonomic computing can be largely divided into two groups: \textit{exogenous} self-ma\-na\-ge\-ment and \textit{endogenous} self-ma\-na\-ge\-ment \cite{weyns2008}. In exogenous self-management the management part is separated from the functional part from the first level of system decomposition (i.e., there is a first level component in charge of the management part) while in endogenous self-management the adaptation at system-level is the result of the collaboration between the system components (i.e., each component fulfills both the functional and the management requirements). We identified three approaches to exogenous self-management: centralized management (e.g., Rainbow \cite{garlan2004}), hierarchical management (e.g., A3 \cite{baresi2011}) and multi-agent management (e.g., self-organizing feedback loops \cite{caprarescu2009}). Centralized management solutions are not suitable for large systems because a central manager introduces a single point of failure and may become a scalability bottleneck. For example, the limitations of centralized solutions were experienced by Meng et al. \cite{meng2010} and were addressed by replicating the central management component. Beyond that, having a single manager in one cloud that supervises components deployed on multiple clouds is like going halfway from single cloud systems to multi-cloud systems. Hierarchical management solutions are appealing in a multi-cloud context, but they raise the additional issues of scaling the managers and finding the right tradeoff between increasing the number of managers to increase the reliability of the system and minimizing the resource consumption of the management part. The same problem characterizes multi-agent solutions as seen in \cite{frincu2011}, where the authors are concerned with the minimization of the number of agents.

As envisioned by Kephart and Chess \cite{kephart2003}, an endogenous autonomic architecture decomposes the system into many interacting autonomic elements where each autonomic element performs both functional and management tasks and the global management is expected to largely arise from the cooperation among autonomic elements. Such a decentralized approach has the advantage that the management part is automatically scaled together with the functional part and benefits from the same level of reliability as the functional part. The disadvantages come from the fact that it is very difficult to adapt complex optimization algorithms which work on a global model (such as queueing or control theory algorithms) to a decentralized environment.

\begin{table}
\begin{center}
    \caption{A survey of several contributions to autonomic resource provisioning}
    \label{table:related-work-survey}
\begin{tabular}{ | p{2.8cm} | p{2.0cm} | p{3.2cm} | p{2cm}|}
	\hline
	\textbf{Contribution} & \textbf{Autonomic architecture} & \textbf{Autonomic mechanism} & \textbf{Provisioning mechanisms} \\
	\hline
	\multicolumn{4}{|c|}{\textit{Customer-centric contributions}} \\
	\hline
	Bonvin et al. \cite{bonvin2011} & Endogenous & Goal (economic model) & Replication, migration\\
	\hline
	EC2 Auto-Scaling \cite{auto-scaling} & Exogenous & Action & Replication \\
	\hline
	Iqbal et al. \cite{iqbal2011} & Exogenous & Action, regression model & Replication \\ 
	\hline
	Meng et al. \cite{meng2010} & Exogenous & Heuristic & Replication \\
	\hline
	Rightscale \cite{rightscale} & Endogenous & Action (voting) & Replication \\
	\hline
	Sharma et al. \cite{sharma2011} & Exogenous & Utility (integer linear programming) & Replication, resizing, migration  \\
	\hline
	Scalarium \cite{scalarium} & Exogenous & Action & Replication \\
	\hline
	Xiong et al. \cite{xiong2011} & Exogenous & Utility (queueing, PI control) & Resizing\\
	\hline
	
	\multicolumn{4}{|c|}{\textit{Provider-centric contributions}} \\
	\hline
	Almeida et al. \cite{almeida2010} & Exogenous & Utility (analytic model) & Resizing  \\
	\hline
	Ghanbari et al. \cite{ghanbari2012} & Exogenous & Utility (control theory) & Resizing  \\
	\hline
	Fox et al. \cite{fox1997} & Exogenous & Action & Replication \\
	\hline
	Wuhib et al. \cite{wuhib2010} & Endogenous & Utility (heuristic) & Replication  \\
		
	\hline
\end{tabular}
\end{center}
\end{table} 

Most of the contributions shown in Table \ref{table:related-work-survey} adopt an exogenous approach to self-management in which the applications are managed by an external service. While the exogenous research solutions are centralized \cite{almeida2010,fox1997,ghanbari2012,iqbal2011,sharma2011,xiong2011}, the exact type of exogenous management (e.g., centralized, hierarchical, multi-agent) offered by commercial solutions \cite{scalarium,ec2} is not publicly available.

Interesting decentralized approaches are the solutions of Bonvin et al. \cite{bonvin2011} and Wuhib et al. \cite{wuhib2010}, as well as the commercial platform Rightscale \cite{rightscale}. Bonvin et al. describe an economic approach for the deployment and dynamic scaling of component-based applications. Each server can host many component instances and is managed by a server agent which decides to migrate, replicate or remove component instances based on their economic balance. The economic balance is computed as a difference between the utility generated by the component from processing requests and the cost of using the hardware resources of the server. Each server agent stores the complete mapping table between the components and the servers that run instances of those components and use a gossip protocol to update this mapping. As each agent has a complete view of the system, although the proposal is decentralized, it is not scalable with respect to the number of components and servers \cite{bonvin2011}.

Another decentralized approach was proposed by Wuhib et al. \cite{wuhib2010}. Their approach is provider-centric and aims to develop a Google Apps Engine-like PaaS for hosting sites in the cloud. An instance of a site is called module and many modules are deployed on one VM. Each VM is managed by a VM manager and the VM managers are connected into a deployment dependent overlay network in which, for each module that is deployed on a VM, the VM manager is directly connected to all VM managers that host instances of the same module. The utility of a module is defined as the ratio between the allocated CPU capacity and the CPU demand, while the utility of the system is defined as the minimum utility of all modules. Authors make a hard assumption with respect to the fact that the CPU resources can be partitioned among the modules running on the same VM. Also, they propose a decentralized heuristic algorithm that maximizes the utility of the system while minimizing the cost of adaptation. The resulting system is scalable with respect to the number of virtual machines and the number of sites, but it is not scalable with respect to the number of modules of a site.

Unlike other commercial solutions, Rightscale uses an endogenous auto-scaling approach in which each VM votes for a scaling action, a decision being taken with the majority of votes \cite{rightscale}. 

The second classification criterion used in Table \ref{table:related-work-survey} is the autonomic mechanism. By autonomic mechanism we refer to the policies and methods used to specify the autonomic behavior and to derive the adaptation actions, respectively.  Kephart and Walsh differentiate among three types of autonomic policies: \textit{action} policies define the specific actions to be taken when the system reaches certain states, \textit{goal} policies divide the states of the system into desirable and undesirable, while \textit{utility} function policies assign a numeric value to each state so that the optimum state to transition can be derived \cite{kephart20}. Although policy-based self-management can be implemented by a rule engine, such as Drools \cite{drools}, which evaluates and executes the policies, goal and utility-based management require more complex optimization methods, such as queuing models \cite{xiong2011}, control theory models \cite{ghanbari2012}, integer linear programming \cite{sharma2011}, economic approaches \cite{bonvin2011}, regression models \cite{iqbal2011}, probabilistic models \cite{almeida2010}, or heuristics \cite{meng2010,wuhib2010}. 
Therefore, most research contributions from Table \ref{table:related-work-survey} use utility or goal-based methods while action polices are largely employed in commercial solutions. An example of action policy is ``if CPU-utilization $>$ \textit{threshold} then add \textit{n} servers of type \textit{T}''.  The problem with action policies is that both the number and type of servers to be provisioned are hardcoded. An improved action-based mechanism is the provisioning method of Meng et al. \cite{meng2010}, which dynamically computes the number of servers to be provisioned based on the throughput increases produced by previous provisioning actions.  

The third classification criterion concerns the provisioning mechanisms, which are in number of three: replication (i.e., scale in/out), resizing (i.e., scale up/down) and migration. As shown in Table \ref{table:related-work-survey}, most solutions are based on replication while there are also a few using resizing. Actually, the focus on replication is justified by the fact that replication is the unique provisioning mechanism offered by some cloud providers like Amazon EC2. The only contribution from Table \ref{table:related-work-survey} where all the three provisioning mechanisms are used is the work of Sharma et al. \cite{sharma2011}, which uses integer linear programming to derive the number and type of VMs to be allocated for each tier of a multi-tier application with the dual goal of satisfying the workload of the tier and minimizing either the cost of resources or the adaptation time. Moreover, they consider two variants of migration: live migration (i.e., moving a component from one machine to another without stopping it) and shutdown-migrate (i.e., the component is stopped, its state is saved and used to start a new instance of that component on another machine). 

Beyond the three aspects discussed above there are two emerging directions in the field of autonomic provisioning for cloud computing that deserve our attention: SLA-aware provisioning and multi-cloud provisioning. Receiving SLA guarantees is an important aspect for the service consumers and this is reflected by the fact that most contributions surveyed in this section use SLA violations (either happened or predicted) to trigger adaptations. More concretely, in all surveyed SLA-aware solutions, the SLA is limited to providing some guarantees with respect to the service response time \cite{bonvin2011,iqbal2011,sharma2011,xiong2011,almeida2010,ghanbari2012}. Thus, if in many papers it is specified that the SLA is met when the response time is kept under a given threshold (e.g., \cite{almeida2010}), in others a utility function is defined to quantitatively asses the SLA fulfillment (e.g., \cite{ghanbari2012}). There are also differences in the way of interpreting the response time: some SLA approaches are based on the average response time (e.g., \cite{xiong2011}),  while others guarantee a maximum response time for every request with a certain probability (e.g., \cite{almeida2010}). However, due to the provisioning latency, it is impossible for a reactive provisioning solution to guarantee a maximum response time for any workload variation. Therefore, some contributions assume the existence of a prediction module capable to provide accurate workload predictions \cite{almeida2010,sharma2011}. 

Sometimes, the offering of high performance and availability service guarantees is constrained by the availability and cost of cloud resources. Therefore, a natural solution for highly available services consists in the ability to provision resources from multiple providers \cite{armbrust2010}. One impediment towards multi-cloud deployments comes from the fact that inter-cloud communication has a higher latency than intra-cloud communication and is usually charged by the providers. This seems to be a tough challenge since no contribution from Table \ref{table:related-work-survey} addresses a multi-cloud scenario, the current multi-cloud research work being limited to a few architectural promises~\cite{buyya2010,celesti2010}. Another multi-cloud impediment consists in the lack of portability and interoperability among different cloud providers and is addressed by the works of Petcu et al. \cite{petcu2011} and Venticinque et al. \cite{venticinque2010}. Thus, on one hand, the vendor-independent API described in \cite{petcu2011} would enable the development of portable applications that can be easily migrated from one cloud to another. On the other hand, the work of Venticinque et al. \cite{venticinque2010} presents the design of a tool for provisioning the best configuration of resources from multiple cloud infrastructures according to a given criterion, such as cost and availability. 

On the industrial side, some cloud computing toolkits such as~\cite{nimbusproject,leadsopennebula,nurmi2009eucalyptus} already offer an abstraction to implement the capability of having different underlying clouds, and auto-scaling/load-balancing over time~\cite{keahey2009sky}. However the solutions offered are centralized or hierarchical, meaning that the decisions to perform auto-scaling and load-balancing actions are made by a centralized controller that is typically offered by the cloud provider. These solutions are well known and used, but may become inadequate in contexts in which the cloud has unreliable nodes, for examples in scenarios in which the cloud resources are offered by small companies or even home users that want to share some unused computational power. In this kind of scenarios an application may experience a continuous unpredictable loss of resources, or, because of Internet connectivity problems, it may also have a consistent instantaneous loss of resources.

In this section, we have surveyed several relevant contributions to the emerging field of auto-scaling in a cloud environment. We have found that the development of robust and scalable provisioning mechanisms for large scale systems remains one of the top challenges in the field together with the challenge of auto-scaling over multiple clouds.
%Therefore we will propose a new approach that holds the promise to address the cloud provisioning problem in a selected context that will be presented in the next section.
%\input{3-context}
\section{\DEPAS{} Approach}
\label{sec:solution}

This section presents \DEPAS{}, a DEcentralized Probabilistic Algorithm for building Auto-Scaling service systems that are capable to cope with high and highly fluctuating request rates.

The goal of our approach is to define an algorithm to support the deployment and the provisioning of batch services in a system composed of federated clouds and an architecture that allows its implementation.
In particular we want to provide a solution that is able to work in an efficient way also in perturbed systems in which the resources are heterogenous in terms of \emph{computational capacity}, defined as the number of operations that a resource is able to execute in a time unit. Beside this we want also a solution that works in situations in which there may be thousands of cheap and unreliable resources that cannot share the full knowledge of all the other resources of the system. It is widely known from the P2P area that using a classical centralized/hierarchical approach in a high-scale system in which elements are unreliable and with local knowledge only may produce high management and coordination costs~\cite{milojicic2002peer}. These costs are due to the rebuilding of the structure of the system in case of failure of the central points. 

%The main contribution of our work consists in a novel auto-scaling method integrated into an architecture for deploying auto-scaling services in the cloud. 

% OVERVIEW OF THE ARCHITECTURE
%Driven by the requirements of highly scalable, multi-cloud service systems, decentralization emerged as the main characteristic of our solution. From the architectural point of view, our solution is a P2P system where each peer represents a resource of the system. Where a resource consists in an autonomic service deployed on a cloud virtual machine. The autonomic services run a probabilistic auto-scaling algorithm in which each autonomic service decides -- independently of other peers -- to add new autonomic services or remove itself in a probabilistic manner. A server-based approach combined with DNS round robin were employed to balance the requests among the autonomic services. 

This section is organized into two subsections. 
Subsection \ref{sub:architecture} shows the architecture of the solution, while the probabilistic auto-scaling algorithm is presented in Subsection \ref{sub:probabilistic-auto-scaling}.
% while two load balancing algorithms that were jointly used in our tests are introduced in Subsection \ref{sub:load-balancing}. Finally, Subsection \ref{sub:overlay-management} describes the gossip protocol used to maintain the P2P overlay network.

\subsection{Architecture to support the auto-scaling service}
\label{sub:architecture}

The proposed architecture is composed of many \textit{autonomic services}. An autonomic service is responsible for processing requests (also called jobs) coming from the clients and, at the same time, for running a process that makes decisions whether to self-destroy or replicate the autonomic service by performing local monitoring activities. Self-destroy and replication actions are performed by exploiting the APIs provided to each autonomic service by the underlying cloud provider(s). In particular, the part of the autonomic service that is responsible for the replication contains the list of the underlying cloud provider(s) and the logic that is needed to choose among them. In our work we are using a random policy for choosing the destination provider for autonomic services replica, leaving the study of more advanced placement heuristics as future work.

\begin{figure}
    \begin{center}
        \includegraphics[scale=0.4]{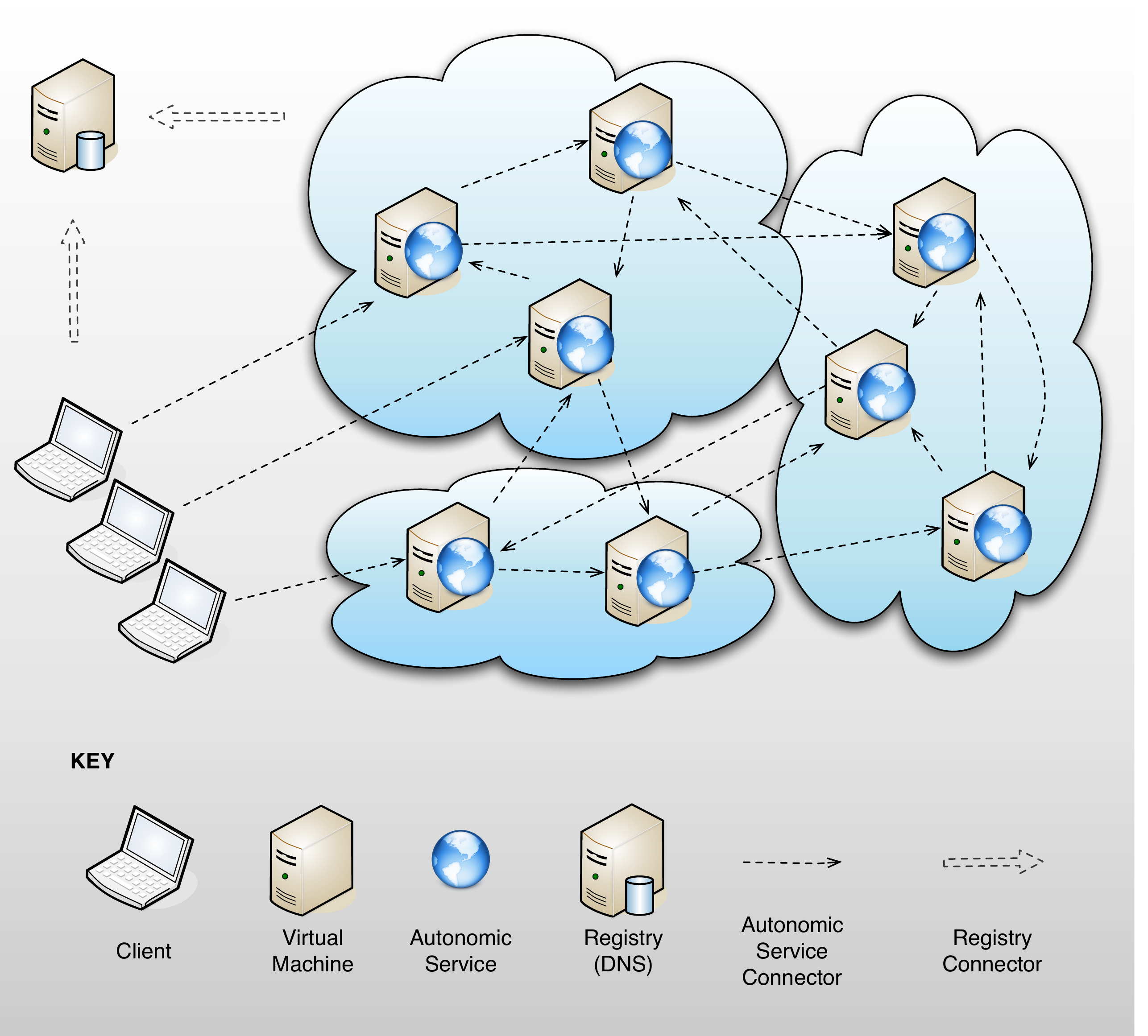}
        \caption{Solution Architecture}
        \label{fig:solution-architecture}
    \end{center}
\end{figure}

%In our architecture, the autonomic element is called autonomic service. It manages several service instances and implements several decentralized algorithms: auto-scaling, load balancing, overlay management, and DNS registration. 
Figure \ref{fig:solution-architecture} shows the architecture that supports the implementation of \DEPAS{}. An autonomic service is deployed on one virtual machine (VM), where a VM represents the resource unit that is offered by the cloud provider.
%As the autonomic service is able to manage many service instances, thus exploiting the full amount of resources provided by the virtual machine, there is no need for deploying more than one autonomic service on the same machine.

Each autonomic service should be able to communicate with other autonomic services (even if they belong to different cloud providers) because the algorithm we will present in the next section requires to diffuse partial information about its state.
Therefore autonomic services are organized into an overlay network in which each autonomic service knows a fixed number of other autonomic services, called \emph{neighbors}. 
% NOT USEFUL INFORMATION
%The links between autonomic services are uni-directional and connectionless. By uni-directional we mean that if service \textit{B} is a neighbor of service \textit{A} then service \textit{A} is not necessarily a neighbor of service \textit{B}. 
% THIS PART SHOULD BE DESCRIBED IN THE SIMULATOR DESCRIPTION SECTION (RESULTS SECTION)
%The autonomic services communicate over UDP and a gossip protocol is used for maintaining the overlay network. The uni-directional UDP connector and the overlay management algorithm allow the auto-scaling operations (i.e., additions and removals of autonomic services) to be performed in a simple, yet effective manner. They also provide the network with built-in fault tolerance in the sense that no special measure needs to be taken when an autonomic service crashes as the other peers progressively remove the links to non-responsive neighbors. 
The \textit{neighborhood} of an autonomic service is composed of the autonomic service itself and its neighbors.

\begin{figure}
    \begin{center}
        \includegraphics[scale=0.4]{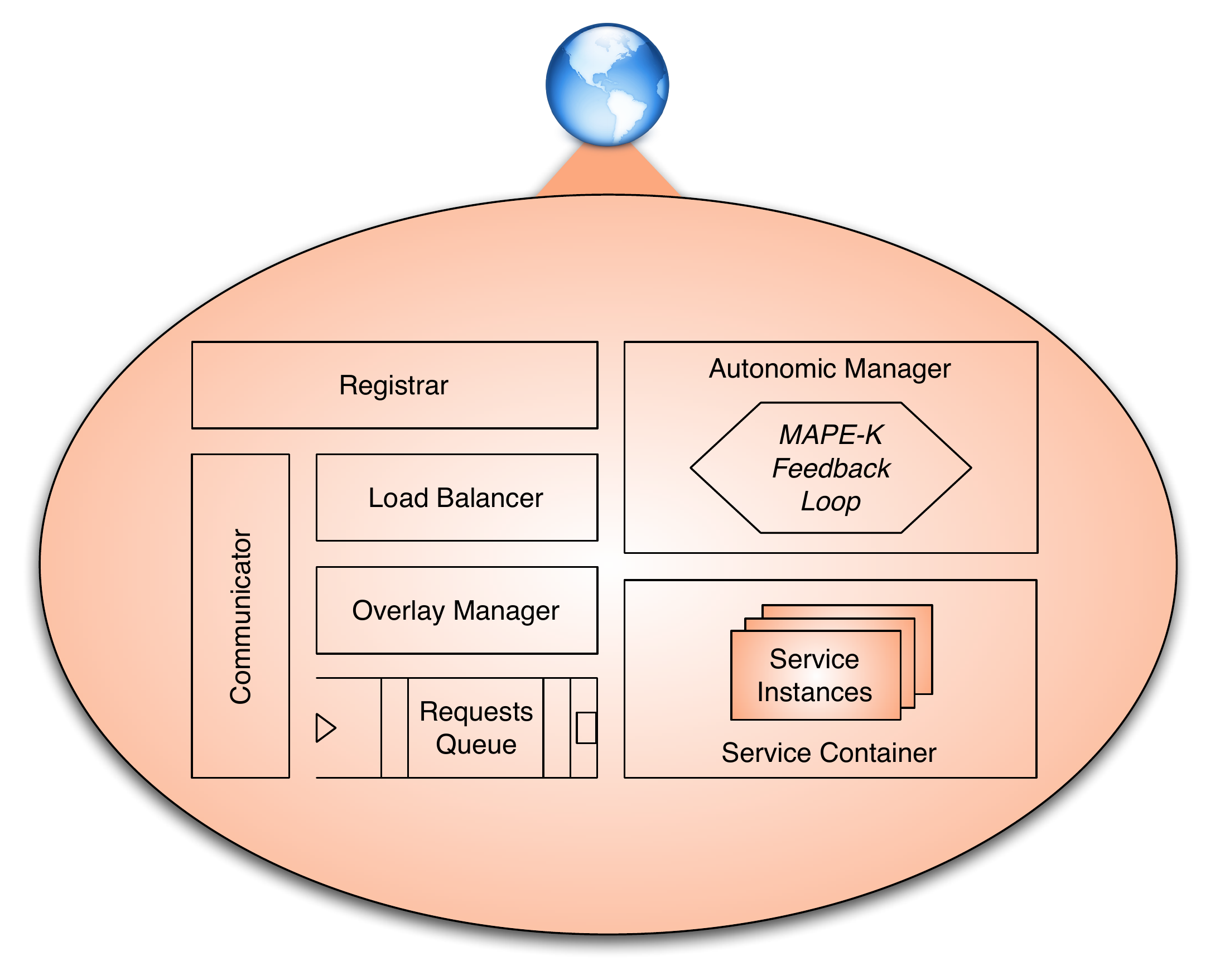}
        \caption{Autonomic Service Architecture}
        \label{fig:autonomic-service-architecture}
    \end{center}
\end{figure}

The internal architecture of the autonomic service is shown in Figure \ref{fig:autonomic-service-architecture}. The \textit{Communicator} allows the autonomic service to communicate with other autonomic services using overlay links.

The \textit{Service Container} instantiates the actual business service that is able to process the requests coming from the clients, and a \textit{Requests Queue} is employed to dispatch the incoming requests to the Service Container. The \textit{Load Balancer} implements a decentralized load balancing algorithm that optimizes the size of the queue (see Subsection~\ref{sub:load-balancing}). A gossip-based algorithm is run by the \textit{Overlay Manager} to maintain the links to the neighbors of the autonomic service (see Subsection~\ref{sub:overlay-management}). The \textit{Registrar} probabilistically decides to register to or unregister from a distributed external registry such as a DNS server.
% and implements the endpoint of the corresponding DNS connector.
This is needed as a mechanism to advertise the existence of the autonomic services that is used by clients for accessing them.

Finally, the \textit{Autonomic Manager} employs a MAPE-K (Monitor-Analyze-Plan-Execute over a Knowledge base) feedback loop \cite{kephart2003} to monitor the load of the current autonomic service and its neighbors, make the provisioning decisions, and execute those decisions. 
\subsection{Probabilistic auto-scaling}
\label{sub:probabilistic-auto-scaling}

In this subsection we describe the auto-scaling logic of the \DEPAS{} approach
%a novel method for service auto-scaling 
that is designed to work with the architecture shown in Subsection \ref{sub:architecture}. The main idea of our method is that each autonomic service decides to create new autonomic services or remove itself in a probabilistic manner and independently of other autonomic services. The purpose of these decisions is to have a total number of autonomic services such that the utilization of each autonomic service stays close to a given threshold.

The auto-scaling algorithm we propose is shown in algorithm \ref{alg:analyze}. In charge of executing it is the \textit{Autonomic Manager} subcomponent of the autonomic service (see the architecture of the autonomic service in Figure \ref{fig:autonomic-service-architecture}). The Autonomic Manager periodically (with period equal to $T^s$) retrieves the neighborhood load (averaged over the last $T^m$ timeframe). If the load is less than the minimum load threshold ($L^{min}$) then the possibility to remove the current autonomic service is considered (see algorithm \ref{alg:analyzeRemoval}). Otherwise, if the average neighborhood load is higher than the maximum load threshold ($L^{max}$) then the autonomic service tries to add new autonomic services (see algorithm \ref{alg:analyzeAddition}).

\begin{algorithm}
\caption{DEPAS}
\label{alg:analyze}
\begin{algorithmic}
\WHILE{$true$} 
\STATE  $wait(T^{s})$
\STATE $L \gets monitor.computeAverageNeighborhoodLoad(T^m)$
\IF {$L < L^{min}$} 
        \STATE $analyzeRemoval(L)$
\ELSE
	\IF {$L > L^{max}$} 
      	\STATE $analyzeAddition(L)$
	\ENDIF
\ENDIF
\ENDWHILE
\end{algorithmic}
\end{algorithm}

Both \textit{analyzeRemoval} and \textit{analyzeAddition} functions rely on the \textit{computeRatio()} function to compute the formula from equation \ref{eq-ratio-homogenous}. This formula is highly parameterized and will be explained in the remainder of this subsection. As this ratio is negative for removals the  \textit{analyzeRemoval} function computes its absolute value which is used as a probability for removing the current autonomic service. In the \textit{analyzeAddition} function, this ratio can be supra-unitary where the integer part represents the number of autonomic services to be added for sure while the fractional part is used as the probability for adding an extra autonomic service.

\begin{algorithm}
\caption{analyzeRemoval(L)}
\label{alg:analyzeRemoval}
\begin{algorithmic}
\STATE $r \gets computeRatio(L)$
\IF {$\mathit{abs}(r) > random()$} 
        \STATE $removeSelf()$
\ENDIF
\end{algorithmic}
\end{algorithm}

\begin{algorithm}
\caption{analyzeAddition(L)}
\label{alg:analyzeAddition}
\begin{algorithmic}
\STATE $r \gets computeRatio(L)$
\STATE $n \gets \lfloor r \rfloor$
\IF {$\left\{r\right\} < random()$} 
        \STATE $n \gets n + 1$
\ENDIF
\STATE $addAdditionalAutonomicServices(n)$
\end{algorithmic}
\end{algorithm}

To derive equation \ref{eq-ratio-homogenous} we have defined a set of  parameters that are summarized in Table \ref{table:notations}. Two important parameters of an autonomic service are the capacity and the load. The capacity (denoted with $C_i$) is the maximum number of requests (jobs) per second that can be processed by the autonomic service $i$ and is derived through benchmarking. 
%The auto-scaling method is formulated for the situation in which all autonomic services have the same capacity $C$. The section ends with the listing of the auto-scaling algorithm.
The load of an autonomic service $i$ (denoted with $L_i(t)$) computed at time $t$ is the ratio between the average number of requests per second that were issued in the time interval $[t - T^m, t]$ and the capacity of the autonomic service. 
%The conditions under which a request is rejected are formulated in Subsection \ref{sub:load-balancing}. By considering both the processed and rejected requests the load of a autonomic service is supra-unitary in cases when the autonomic service receives more requests than it can process. 
All the load parameters, such as the average load per autonomic service of the system and the load thresholds are also expressed as percentages with respect to the capacity of the autonomic service.

\begin{table}
\begin{center}
    \caption{Auto-scaling notations}
    \label{table:notations}
\begin{tabular}{ | p{1.8cm} | p{9cm} | }
	\hline
%	$\lambda(t)$ & Mean request rate at a given time (number of requests per second) \\
%	$\mu$ & Mean service time (in seconds) \\
	$N(t)$ & Actual number of autonomic services allocated at a given time \\
	$\hat{N}(t)$ & Target number of autonomic services to be allocated at a given time \\
	$M(t)$ & Number of autonomic services to be added/removed at a given time \\
	$C_i$ & Capacity of autonomic service $i$\\
	$C^{\mathit{av}}$ & Average capacity among all system autonomic services\\
	$L_i(t)$ & Load of autonomic service $i$ computed at time $t$ over timeframe $(t-T^m,t)$ (percent with respect to autonomic service capacity) \\
	$L^{\mathit{av}}(t)$ & Average load per autonomic service of the system, computed at time $t$ over timeframe $(t-T^m,t)$ for all autonomic services in the system (percent) \\
	$\tilde{L}^{\mathit{av}}_{i}(t)$ & Average load per autonomic service of the neighborhood of autonomic service $i$, computed at time $t$ over timeframe $(t-T^m,t)$ (percent)  \\
	$L^{min}$ & Minimum load threshold  (percent) \\
	$L^{max}$ & Maximum load threshold  (percent) \\
	$L^{des}$ & Desired load threshold, which is equal to $(L^{max}-L^{min})/2$  (percent) \\
	$T^s$ & Period between two successive runnings of the auto-scaling algorithm on a autonomic service \\
	$T^m$ & Length of monitoring timeframe for the actual load \\
	$\mathit{neighborhood}_i$ & Neighborhood of autonomic service $i$: contains autonomic service $i$ and its neighbors \\
	$\mathit{queue}_i$ & Number of enqueued requests in autonomic service $i$ \\
	$R_{\mathit{max}}$ & Maximum response time for completed requests (from SLA) \\
	\hline
\end{tabular}
\end{center}

\end{table}

Given at time $t$ the actual number of nodes of the system $N(t)$, the  average load $L^{\mathit{av}}(t)$, a desired target load $L^{\mathit{des}}$, and an average capacity $C^{\mathit{av}}$ that is always constant independently from the number of the nodes, then the target number of nodes at time $t$ is calculated as follows:

\begin{equation}
\label{eq-N}
\hat{N}(t) = \frac{L^{\mathit{av}}(t)}{L^{\mathit{des}}} \cdot N(t)
\end{equation}

The explanation for equation \ref{eq-N} is that we want to redistribute the total load of the system (defined as $L^{\mathit{av}}(t) \cdot N(t)$) in a system in which the autonomic services have different weights in terms of load (from $L^{\mathit{av}}(t)$ to $L^{\mathit{des}}$).

Therefore the total number of autonomic services to be added (positive value) or removed (negative value) from the system is the following:

\begin{equation}
\label{eq-M}
M(t) = \hat{N}(t) - N(t) =   \frac{L^{\mathit{av}}(t) - L^{\mathit{des}}}{L^{\mathit{des}}} \cdot N(t)
\end{equation}

Computing $M(t)$ in presence of global information such as $N(t)$ and $ L^{av}(t)$ is very simple, but we aim to provide a decentralized solution in which each autonomic service decides to add new autonomic services or remove itself in a probabilistic way. We argue that the ratio $R_{i}(t) = M(t)/N(t)$ can be used for deriving this probability. By making these simplifications equation \ref{eq-mn-ratio} is obtained:

\begin{equation}
\label{eq-mn-ratio}
R_{i}(t) =  \frac{M(t)}{N(t)} =  \frac{L^{\mathit{av}}(t) - L^{\mathit{des}}}{L^{\mathit{des}}}
\end{equation}

Considering that we opted for a P2P architecture in which each autonomic service is aware of a small subset of all autonomic services, computing the average load of the system at each autonomic service is not feasible, but there are a few solutions for approximating this value. In our simulator, each autonomic service uses the weighted average load of its neighborhood as an approximation of the average load of the system using capacities as weights (see equation \ref{eq-av-load-homogenous}). 

\begin{equation}
\label{eq-av-load-homogenous}
\tilde{L}^{\mathit{av}}_{i}(t) =  \frac{\sum_{j \in \mathit{neighborhood}_i} (C_j \cdot L_j(t))}{\sum_{j \in \mathit{neighborhood}_i} C_j}
\end{equation}

Then, this value can be used to calculate a local estimator for the ratio $R_{i}(t_1)$, that we have denoted in equation~\ref{eq-ratio-homogenous} as $\tilde{R}_{i}(t_1)$ and used in algorithms \ref{alg:analyzeRemoval} and \ref{alg:analyzeAddition} as the \emph{ComputeRatio()} function.

\begin{equation}
\label{eq-ratio-homogenous}
\tilde{R}_{i}(t) =  \frac{\tilde{L}^{\mathit{av}}_{i}(t) - L^{\mathit{des}}}{L^{\mathit{des}}}
\end{equation}

%In the case of scaling in, the ratio $\tilde{R}_{i}(t_1)$ is a negative (between -1 and 0), while in the case of scaling out it is positive (greater then 0). 
When $\tilde{R}_{i}(t)<0$, autonomic service $i$ uses $|\tilde{R}_{i}(t)|$   as a probability to remove itself. When $\tilde{R}_{i}(t)>0$, the fractional part of $\tilde{R}_{i}(t)$ is used as a probability to add a new autonomic service, while the integer part (which is greater than zero when the number of autonomic services to be added is higher than the number of existing autonomic services) is the number of autonomic services that will be added in a deterministic way.
The capacity of the new autonomic service is assumed to be, in the average case, approximated to the average capacity in the neighborhood of the local autonomic service.

In addition to the decision rules for adding and removing resources to the system, our probabilistic auto-scaling approach requires the adoption of proper fault-tolerant mechanisms to keep the load among the autonomic services balanced and the topology connected. For this purpose we have adopted and customized some existing approaches that will be described in the following two paragraphs.

\subsubsection{Supporting Auto-scaling with Load-balancing}
\label{sub:load-balancing}

The autoscaling algorithm described above needs to be supported by a load balancing algorithm because otherwise the new autonomic services added by the provisioning algorithm would never receive the existing load of the system.
%provisioning of new autonomic services is useless unless the incoming traffic is dynamically load balanced among all autonomic services. 
We opted for a two-level load balancing: a DNS-based round-robin entry point combined with a decentralized approach at autonomic service level. At DNS level a configurable percentage of all autonomic services are registered in the external DNS which offers one autonomic service from its list to the clients in a round-robin fashion. At autonomic service level we tested a combination of two load balancing algorithms: \textit{admission control} and \textit{dimension exchange}.

Admission control is the name we give to the load balancing algorithm used in \cite{adam2006}. The idea is that, when an autonomic service receives a request, an admission function is used to decide whether the request will be processed by the current autonomic service, forwarded to a remote autonomic service, or rejected. 
%In \cite{adam2006} the admission control is done by estimating the response time of the request and comparing it with a maximum response time threshold. The estimated response time is computed as the sum between the time that has been already spent while routing the request, the foreseen time needed to process the requests that are already in the queue, and the foreseen execution time of the request itself. If the admission control fails then the request is forwarded to one neighbor in the overlay. A list of visited autonomic services is stored in the request header and used to prevent the request from visiting twice the same autonomic service. Eventually the request is rejected if it cannot be processed within the maximum response time even by a free autonomic service (i.e., an autonomic service with an empty queue).

%Actually, various policies can be employed for the admission function, neighbor selection, and request rejection. Thus, we opted for an admission policy that is independent of response time because of the difficulty of deriving accurate time predictions. More concretely, we adopt a limit in the number of hops traversed by the request, and two queue size limits: a \emph{soft} limit and a \emph{hard} limit. 
We have chosen an admission policy that uses three limits: a forwarding limit, a soft queue limit, and a hard queue limit.
If the length of the queue is less than the soft limit, or the request has already been forwarded the maximum number of times and at the same time the queue is less than the hard limit, then the request is admitted and enqueued in the local autonomic service. If the length of the queue is equal or higher than the soft limit, and the request has not reached yet the maximum number of forwardings, then the request is forwarded to a random autonomic service that has not already been visited by the same request. In other cases the request is rejected.
%the admission control passes if the length of the queue is less than a given threshold. 
%We use a ponderal random strategy for neighbor selection in which the neighbors capacities act as ponders and the request is rejected when the number of visited autonomic services reached a given threshold. 
%The resulting algorithm is shown in listing \ref{alg:admission-control}.

%\begin{algorithm}
%\caption{Admission control load balancing}
%\label{alg:admission-control}
%\begin{algorithmic}
%\WHILE{$true$} 
%\STATE $r \gets receiveRequest()$
%\STATE $r.visit(this)$
%
%\IF {$queue.size <  \mathit{SoftQueueLimit}$} 
%	\STATE $queue.append(r)$
%\ELSIF {$r.getNbOfHops() \leq \mathit{MaxNbFwds}$} 
%	\STATE $n \gets selectNeighbor()$
%	\STATE $send(r, n)$		
%\ELSIF {$queue.size <  \mathit{HardQueueLimit}$} 
%	\STATE $queue.append(r)$
%\ELSE
%	\STATE $reject(r)$
%\ENDIF
%\ENDWHILE
%\end{algorithmic}
%\end{algorithm}

The second load balancing algorithm used in the simulator is an adapted version of the dimension exchange algorithm we also used in our past work~\cite{dinitto2008}. The idea of the original algorithm is that two autonomic services \textit{A} and \textit{B} cooperate to average their requests. We have modified the formula of our previous approach by inserting also terms for autonomic service capacities that are used as weights to compute the weighted average. %Requests can be moved either from \textit{A} to \textit{B} or from \textit{B} to \textit{A}. The problem with this algorithm is that it requires the implementation of a timeout-based failure detection that would allow one autonomic service to cancel the algorithm in the situation the other autonomic service fails. Rather than implementing such fault tolerance mechanism we preferred to simplify the algorithm by allowing the requests transfer only from the passive autonomic service to the active autonomic service. For example, suppose that autonomic service \textit{A} sends the size of its queue to neighbor \textit{B}.  Autonomic service \textit{B}, upon receiving the queue size of \textit{A}, computes the number of requests to be transferred from \textit{B} to \textit{A} using the formula from equation (\ref{eq-transfer}). If $T(B, A) > 1$ then $T(B, A)$ requests are moved from the queue of A to the queue of B. 
Therefore the number of requests moved from $A$ to $B$ ($B$ to $A$) is calculated by $T(A,B)$ ($T(B,A)$) using equation \ref{eq-transfer}.

\begin{equation}
\label{eq-transfer}
T(A, B) = -T(B, A) = \left\lfloor \frac{|\mathit{queue}_A| \cdot C_B-|\mathit{queue}_B| 
 \cdot C_A}{C_A+C_B}\right\rfloor
\end{equation}

%In both admission control and dimension exchange load balancing losing messages that transfer requests from one autonomic service to another is not acceptable because it would negatively impact the QoS of the service. Therefore, as stated in Subsection \ref{sub:architecture} a reliable UDP connector is used for this purpose. The reliable UDP connector informs the sender when the message delivery failed, thus allowing the sender to forward the message to another neighbor and remove the link to the failed neighbor. Anyway, as described in Subsection \ref{sub:overlay-management}, the dead link would have been eventually removed by the build-in fault tolerance mechanism of the overlay management protocol. 

In both algorithms when the permanence of the requests in the queue becomes higher than $R_\mathit{max}$ (the maximum response time of a request before being rejected that can be found in the SLA), then the request is removed from the queue and rejected.

\subsubsection{Supporting Auto-scaling with Overlay Management}
\label{sub:overlay-management}

At the beginning of this section we have justified the need of an overlay network that interconnects the autonomic services to support the information exchanges needed by the auto-scaling algorithm. Since our solution needs to operate also in presence of dynamism it is important that the system overlay is maintained connected and with a target degree (where the overlay degree is defined as the average number of neighbors per autonomic service) even in situations in which the appearance/disappearance of autonomic services is frequent and unpredictable. For this reason we decided to support the overlay with an overlay management protocol.
The topology management protocol we have adopted is an adapted version of the gossip protocol developed by Jelasity et al.~\cite{Jelasity2007}.
%The main requirements of the overlay management protocol can be derived from the architectural requirements highlighted in Subsection \ref{sub:architecture}. 
The motivation for adopting this protocol is that it has been already proven that it is  highly scalable, highly reliable with respect to autonomic service failures, and the operations of neighbor addition and removal required by our auto-scaling algorithm can be done with minimum effort. 
% THIS PART SHOULD BE MOVED TO THE STATE OF THE ART
%Gossip protocols are appealing in building large-scale unstructured overlay networks that are highly resilient to failures and high churn rates \cite{Jelasity2007}.  In \cite{Jelasity2007}, Jelasity et al. describe a generic protocol that can be customized to create a large spectrum of concrete gossip protocols. 

The overlay management protocol we used maintains a uni-directional overlay network where each autonomic service keeps a list with a fixed number of autonomic service entries (called \textit{neighbors view}). At the end of each time frame $T_\mathit{overlay}$ a protocol iteration is executed in each autonomic service (autonomic services are not synchronized). After each iteration the neighborhood of an autonomic service is merged with the neighborhood of a neighbor autonomic service.
The fixed number of neighbors of any autonomic service is called the degree of the network and noted with $c$, while parameters $H$ and $S$ of the protocol are used to determine the fault-tolerance and persistence of the neighbors in the local view. Additional details on these parameters is explained in~\cite{Jelasity2007}.

An issue with the original Jelasity protocol is that the actual overlay degree remains equal to $c$ even in the situation when the total number of autonomic services is less than $c$, moreover it is possible that some old autonomic services that have been removed from the network are never removed from the neighborhoods of existing autonomic services. To solve it we have used a customized variant that introduces a maximum age parameter, $o$: at the beginning of the \textit{merge} operation the entries with an age higher than $o$ are removed from both the neighbors views. A proper value for $o$ depends on the values of the others parameters of the overlay and can be found through simulations.
%The parameters of our protocol are summarized in Table \ref{table:our-overlay}.

%The resulting overlay management algorithm used in our simulator is described in listings \ref{alg:active-node}, \ref{alg:passive-node}, and \ref{alg:view.merge}. It represents both a specialization and an adaptation of the Jelasity generic gossip protocol \cite{Jelasity2007}. We do not argue that it is generally better than other protocols but it fulfills our requirements.

%\begin{algorithm}
%\caption{active node}
%\label{alg:active-node}
%\begin{algorithmic}
%\WHILE{$true$} 
%\STATE  $wait(T^{overlay})$
%\STATE $n \gets view.selectNode()$
%\STATE $myLoad \gets monitor.getMyLoad()$
%\STATE $exchangeView = ((myAdress, 0, myLoad, myCapacity))$
%\STATE $exchangeView.append(view.head(c/2 - 1))$
%\STATE $send(exchangeView, n)$
%\STATE $myView.increaseAge()$
%\ENDWHILE
%\end{algorithmic}
%\end{algorithm}

%\begin{algorithm}
%\caption{passive node}
%\label{alg:passive-node}
%\begin{algorithmic}
%\WHILE{$true$} 
%\STATE $receive(exchangeView)$
%\STATE $myView.merge(exchangeView, c, o)$
%\ENDWHILE
%\end{algorithmic}
%\end{algorithm}

%\begin{algorithm}
%\caption{view.merge(exchangeView, c, o)}
%\label{alg:view.merge}
%\begin{algorithmic}
%\STATE $removeOldEntries(o)$
%\STATE $exchangeView.removeOldEntries(o)$
%\STATE $exchangeView.removeSelfEntry()$
%\STATE $append(exchangeView)$
%\STATE $removeDuplicates()$
%\STATE $sort()$
%\STATE $removeTail(size() - c)$
%\end{algorithmic}
%\end{algorithm}

\section{Results}
\label{sec:results}

In this section we provide a description of the results obtained by our technique and the methodology we adopted to obtain them.
We stressed our system with realistic input, i.e. with a high variability workload with peaks of usage inspired by a real scenario (see Subsection~\ref{sub:experiments}).
The general approach we adopt is to feed some input (i.e., service requests) to the system and monitor its reaction with respect to different factors (i.e., response time, number of autonomic services instantiated, rejected requests, etc.). The experimental evaluation must consider various aspects which are important either for the client (i.e., response time, rejected/lost requests) or the cloud provider which is interested in reducing costs (i.e., number of autonomic services instantiated).

In Subsection \ref{sub:setting} we give a description of the various settings employed in the experiments, then the actual results are presented in Subsection \ref{sub:experiments} and some findings are presented in Subsection \ref{sub:discussion}.

We planned the experimental phase in order to verify some hypothesis about the behavior of our technique. Specifically we are interested in assessing the 
(i) scalability of the approach with respect to high variations in the input workload and, as a consequence, the presence of a large number of instantiated autonomic services, 
(ii) stability and responsiveness of the response time experienced by the final user, 
(iii) optimality of the approach with respect to a centralized system taking advantage of a global common knowledge and always taking the best action,
%(this aspect is particularly important as we can evaluate how much our approach is over allocating autonomic services that represents direct costs for the service provider), 
and finally (iv) the behavior of the approach in presence of dynamism in the infrastructure such as autonomic services leaving/joining the system or severe damages disrupting a large part of the infrastructure.

\subsection{Experimental setting}
\label{sub:setting}

Considering the large scale nature of our technique, involving potentially thousands of autonomic services, we decided to validate it through a simulation environment implemented on top of the Protopeer platform \cite{protopeer}.
The reason for this choice is that Protopeer allows to run simulations of P2P algorithms over a time domain, even if it is event-based. This is possible thanks to the possibility to schedule events using our time unit of choice (second).
Another interesting feature of Protopeer is that the same simulation code can run all simulated peers in a single machine using a special scheduler with a simulated clock, or one peer per machine using using the real (synchronized) clock of the involved machines. When carrying out our experiments we have chosen to run the simulator in single machine mode because we needed to scale up to 10.000 peers (each one containing one autonomic service).

%Protopeer allows to write prototypes of peer-to-peer applications and to run them in two different settings: 
%(i) \emph{event-based} simulation, the behaviour of the simulation is defined by a set of events and the reactions associated upon the event reception and (ii) \emph{live deployment} in which the same code developed for the event-based system can be deployed in a real computer network. 
%We decided to adopt the event-based simulation as it easily allowed us to prove the stability of our solution in a controlled scenario involving a large number of autonomic services (up to 10.000). Live deployment experiments add a new level of complexity that we plan to study in future analysis.

In order to reduce possible effects due to algorithm dependent random factors we performed a series of 32 independent runs for each experimental configuration; plots include average, min and max values for each one of the measured values. Experiments have been performed on top of 14 virtual machines equipped with 68.4 GB of memory and 8 virtual cores using the Amazon EC2 cloud service \cite{amazon_ec2}.

We conducted experiments in various scenarios varying different aspects such as computing infrastructure parameters (i.e., distribution of capacities to reproduce a situation of multiple cloud providers), submitted requests (i.e., workload) and presence of external system factors (i.e., failures) \footnote{Our experiments can be downloaded at \url{http://home.dei.polimi.it/calcavecchia/depas/depas.zip}}.

% maybe we can also include the impact of load balancing and admission control
In Table \ref{table:fixed_experiment_parameters} we report the values of parameters that are fixed over all the experiments; the values have been obtained by empirical observations of dedicated experiments and show a good tradeoff between the quality of reached solution (i.e., response time optimality, reactivity, etc.) and the cost involved in computing it (i.e., number of messages sent).

\begin{comment}
	\hline
	$S$ & Swap & 0 \\
	\hline
	$H$ & Heal & 15 \\
\end{comment}

\begin{table}
\begin{center}
    \caption{Fixed experiment parameters}
    \label{table:fixed_experiment_parameters}
\begin{tabular}{ | c | p{5cm} | c | }
	\hline
	\textbf{Symbol} & \textbf{Description} & \textbf{Value} \\
	\hline
	$c$  & Overlay degree & 60 \\
	\hline
	$T^{overlay}$ & Period between two protocol iterations & 0.3 sec\\ 
	\hline
	$o$ & Maximum entry age in neighborhood view & 30 \\
	\hline
	$\mu$ & Average service execution time & 1 sec\\ 
	\hline
	$L^{min}$ & Minimum load & 0.6 \% \\
	\hline
	$L^{max}$ & Maximum load & 0.8 \%\\
	\hline
	$L^{des}$ & Desired load & 0.7 \% \\
	\hline
	$QL^{max}$ & Maximum queue length & 20 \\
	\hline
	$PT^{max}$ & Maximum request pending time & 4 sec \\
	\hline
	$DNS_{entries}$ & Number of entries in the DNS & 30 \\
	\hline	
\end{tabular}
\end{center}
\end{table} 

For simplicity we consider a single client issuing service requests. Service execution time is exponentially distributed with mean $\mu=1$.

%Each experiments is executed for 45 minutes 
%with a client issuing service requests. 

In order to stress our system with a realistic workload we defined a service request track having the same pattern found in an on-line collaborative service during the first Town Hall meeting given by the US president Obama in December 2008 \cite{town_hall}. The service was monitored for a 48-hours period with a maximum load of 700 hits-per-second and about 92 thousands users. 
In our experiments we maintained the same request pattern while scaling up the amount of requests per second by a factor of 10 with the objective of producing higher load on the system and consequently a higher number of autonomic services instantiated (the system scales up to 10.000 services in the maximum peak). In order to challenge our system with a highly variating workload, we scaled down the time axis to 45 minutes. In this way, decisions must be taken even more quickly than the case of 48 hours. Finally, requests are distributed according to an exponential distribution with average defined by the previously described request track.

\subsection{Experimental results}
\label{sub:experiments}
We start the presentation of experimental results by first introducing a ``reference experiment'' from which all other experiments are then derived. This first experiment is also used to introduce the main metrics adopted and how we have computed them. In the reference experiment, we assign capacities according to the following probabilistic distribution: 50 \% of autonomic services with capacity 0.5 req/sec, 30 \% with capacity 1.83 req/sec and 20 \% with capacity 1 req/sec.

Figure \ref{fig:referece_requests_cumulative} shows a plot of requests over all system (we call it ``cumulative''). The issued requests line represents the total number of requests that are submitted to our system while the processed requests line shows the throughput of the system. In all the plots the line is the result of an average performed over 32 experiments while the shade around the line represents the range area delimited by minimum and maximum value over the experiments.
The plot shows how the system is able to quickly adapt itself providing a throughput very close to the required one still maintaining a limited number of rejected and lost requests (requests are rejected when they cannot be processed within the maximum response time and requests are lost when an autonomic service in charge of executing them faces a failure).

\pgfplotsset{tick label style={
	font=\footnotesize},
    label style={font=\small},
    legend style={font=\small}
}
\pgfplotsset{compat=1.3}

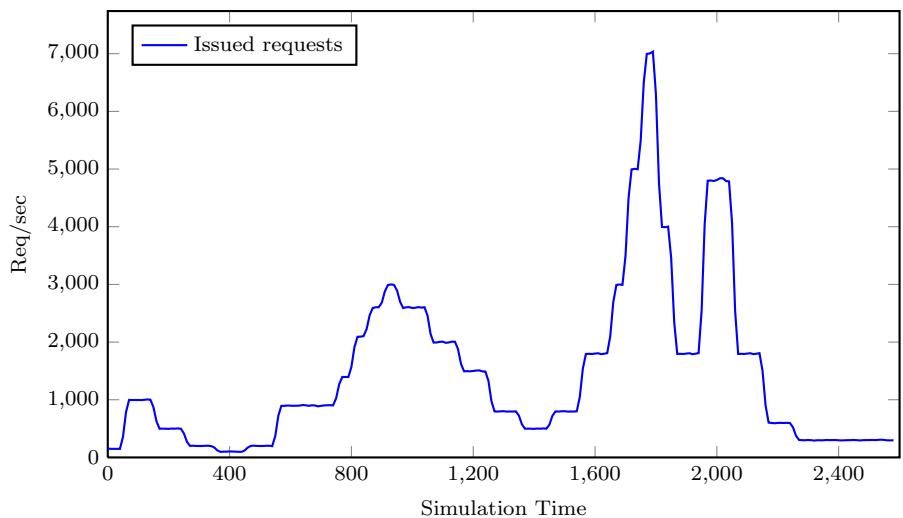
\begin{figure}
    \centering
    \begin{tikzpicture}
        \begin{axis}[
        small,
        samples=200,
		width=12cm,
		height=7.5cm,
		xlabel=Simulation Time,
		ylabel=Req/sec,
		xmin=0,xmax=2600,ymin=0,
		xtick={0, 400,800,1200,1600,2000,2400,2800},
		cycle list={
			{blue},
			{red},
			{orange},
			{cyan,dashed},
			{gray,dashed}},
		thick,
		legend pos= north west,
		legend columns=2,
		legend style={/tikz/every even column/.append style={column sep=0cm, row sep=0cm}}]
		]
			
		\addplot table[
		x=Time,
		y=CLIENT_NB_OF_ISSUED_REQ_Avg
		]{reference.dat};

        \legend{{\footnotesize Issued requests}}
        
        \end{axis}
    \end{tikzpicture}
    \caption{Cumulative requests in reference experiment}
    \label{fig:referece_requests_cumulative}
\end{figure}

%\begin{figure}[htbp]
%    \begin{center}
%			\includegraphics[scale=0.48]{charts/reference_experiment/requests_cumulative} 
%        \caption{Cumulative requests in reference experiment}
%        \label{fig:referece_requests_cumulative}
%    \end{center}
%\end{figure}

In Figure \ref{fig:reference_number_of_nodes} we report the number of autonomic services instantiated in the system at each time instant together with the number that would have been instantiated by an optimal controller in case of desired, minimum and maximum load.
The optimal number of autonomic services is computed according to Formula \ref{for:optimal_nodes} where $C^\mathit{avg}$ is the average capacity of services in the system and $L$ is the load to impose on services (i.e., $L^{min}$, $L^{des}$ and $L^{max}$). 

\begin{equation}
\label{for:optimal_nodes}
N_{\mathit{opt}}(t) = \frac{\lambda (t)}{C^\mathit{avg} * L} 
\end{equation}

Notice that in a centralized system we can assume to know the real value of $C^\mathit{avg}$ thanks to a global knowledge. As the plot shows, the number of autonomic services instantiated in the system closely follows the trend of the desired optimal one either in high and low periods. Phenomena of temporary over provisioning or under provisioning are limited to some very rare cases.

\pgfplotsset{tick label style={
	font=\footnotesize},
    label style={font=\small},
    legend style={font=\small}
}
\pgfplotsset{compat=1.3}

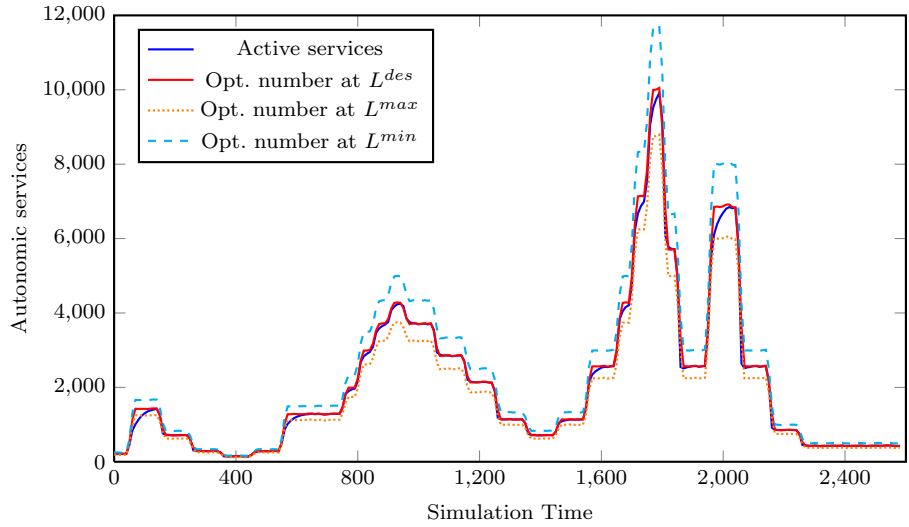
\begin{figure}
    \centering
    \begin{tikzpicture}
        \begin{axis}[
        	scaled y ticks = false,
     	y tick label style={/pgf/number format/fixed},
        small,
		width=12cm,
		height=7.5cm,
		xlabel=Simulation Time,
		ylabel=Autonomic services,
		xmin=0,xmax=2600,ymin=0,ymax=12000,
		xtick={0, 400,800,1200,1600,2000,2400,2800},
		cycle list={
			{blue},
			{red},
			{orange,densely dotted},
			{cyan,dashed},
			{gray,dotted}},
		thick,
		legend pos= north west,
		legend columns=1,
		legend style={/tikz/every even column/.append style={column sep=0cm, row sep=0cm}}]
		]
			
		\addplot table[
		x=Time,
		y=PEER_NB_OF_ACTIVE_WORKERS_Avg
		]{reference.dat};
		
		\addplot table[
		x=Time,
		y=OPTIMAL_NUMBER_AT_DESIRED_LOAD_Avg
		]{reference.dat};

		\addplot table[
		x=Time,
		y=OPTIMAL_NUMBER_AT_MAX_LOAD_Avg
		]{reference.dat};

		\addplot table[
		x=Time,
		y=OPTIMAL_NUMBER_AT_MIN_LOAD_Avg
		]{reference.dat};

        \legend{{\footnotesize Active services}, 
        {\footnotesize Opt. number at $L^{des}$}, 
        {\footnotesize Opt. number at $L^{max}$},
        {\footnotesize Opt. number at $L^{min}$}}
        
        \end{axis}
    \end{tikzpicture}
    \caption{Number of autonomic services instantiated in reference experiment}
    \label{fig:reference_number_of_nodes}
\end{figure}
%\begin{figure}[htbp]
%    \begin{center}
%			\includegraphics[scale=0.48]{charts/reference_experiment/number_of_nodes} 
%        \caption{Number of autonomic services instantiated in reference experiment}
%        \label{fig:reference_number_of_nodes}
%    \end{center}
%\end{figure}

Another perspective on the same data is given by the plot in Figure \ref{fig:referece_requests_peer} where received requests and processed requests at peer level are shown (averaged over all peers). In a perfectly balanced ideal system each autonomic service would receive and process the same amount of load (in percent); from the plot we see that the general trend of received load is constant around the $L^{\mathit{des}}$ value with some spikes. In particular upward spikes happen in correspondence of a sudden increase in load (i.e., the system is not immediately able to instantiate new autonomic services). Conversely, downward spikes are associated with sudden decrease of load which is not instantaneously reflected in decrease of services.

\pgfplotsset{tick label style={
	font=\footnotesize},
    label style={font=\small},
    legend style={font=\small}
}
\pgfplotsset{compat=1.3}

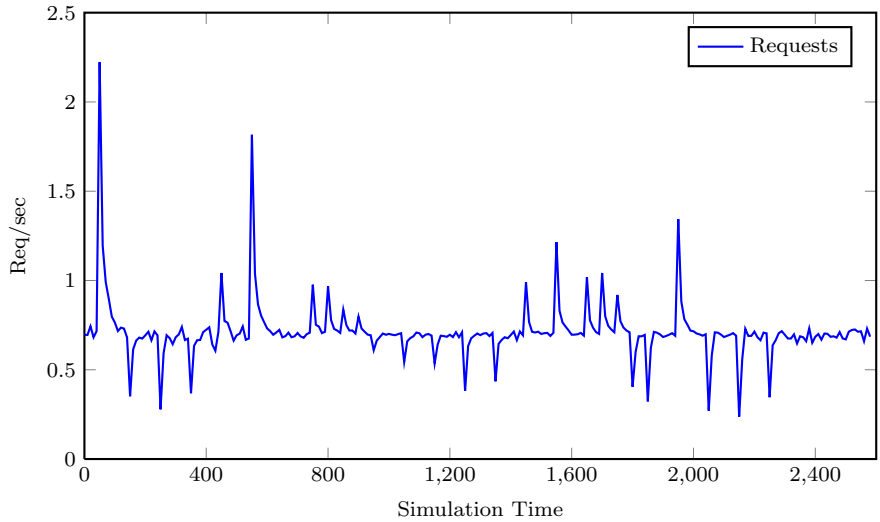
\begin{figure}
    \centering
    \begin{tikzpicture}
        \begin{axis}[
        	scaled y ticks = false,
     	y tick label style={/pgf/number format/fixed},
        small,
		width=12cm,
		height=7.5cm,
		xlabel=Simulation Time,
		ylabel=Req/sec,
		xmin=0,xmax=2600,ymin=0,ymax=2.5,
		xtick={0, 400,800,1200,1600,2000,2400,2800},
		cycle list={
			{blue},
			{red},
			{orange,densely dotted},
			{cyan,dashed},
			{gray,dotted}},
		thick,
		legend pos= north east,
		legend columns=1,
		legend style={/tikz/every even column/.append style={column sep=0cm, row sep=0cm}}]
		]
			
		\addplot table[
		x=Time,
		y=PEER_LOAD_Avg
		]{reference.dat};
		
        \legend{
        {\footnotesize Requests}}
        
        \end{axis}
    \end{tikzpicture}
        \caption{Average requests in reference experiment at autonomic service level}
        \label{fig:referece_requests_peer}
\end{figure}
%\begin{figure}
%    \begin{center}	
%			\includegraphics[scale=0.48]{charts/reference_experiment/requests_peer_level} 
%        \caption{Peer level requests in reference experiment}
%        \label{fig:referece_requests_peer}
%    \end{center}
%\end{figure}

The final measure that the client perceives in the system is the response time of a request (i.e., the difference between the moment the request is received back from the client and the moment in which the request is issued). Figure \ref{fig:reference_response_time} shows the average response time over all requests for each time instant together with the \emph{optimal response time} and the \emph{pending time}. The optimal response time is the one that an ideal centralized system would achieve. More precisely, we assume that in the optimal system there is a central component with a global knowledge about the state of autonomic services (i.e., requests in the queue, capacity, etc.). 
The centralized system always makes the best decision by forwarding the request to the best available autonomic service (i.e., the one that guarantees minimum response time). 
We modeled the centralized system using a M/M/m queue where $m$ is the number of currently  instantiated autonomic services and we computed the response time obtained in this system using well known formulas of the literature \cite{bolch}. Since the centralized system is still distributed, with a client issuing requests to a server over a network infrastructure, we must also consider the network latency in the optimal system too (introduced artificially in the simulator); to this end, we added twice (round-trip) the value of the average network latency to the previously computed value.

As can be seen from Figure~\ref{fig:reference_response_time} the obtained response time remains close to the optimal one with an average value of 1.23 seconds. The response time reflects the sudden changes in the input workload (i.e., sec 100, 600, 1600, 2000). However, it can be observed that the system quickly readapts itself (i.e., instantiating an appropriate number of autonomic services), thus bringing the response time close to the average. The plot also shows the pending time, which represents the time that a request spends in the queue before being processed. 

%\documentclass[a4paper]{article}

%\usepackage{pgfplots}
%\pgfplotsset{compat=1.3}

%\begin{document}

\pgfplotsset{tick label style={
	font=\footnotesize},
    label style={font=\small},
    legend style={font=\small}
}
\pgfplotsset{compat=1.3}

\begin{figure}
    \centering
    \begin{tikzpicture}
        \begin{axis}[
        small,
        samples=200,
		width=12cm,
		height=7.5cm,
		xlabel=Simulation Time,
		ylabel=Seconds,
		xmin=0,xmax=2600,ymin=0,ymax=4,
		xtick={0, 400,800,1200,1600,2000,2400,2800},
		cycle list={
			{blue},
			{red},
			{orange},
			{cyan,dashed},
			{gray,dashed}},
		thick,
		legend columns=2,
		legend style={/tikz/every even column/.append style={column sep=0cm, row sep=0cm}}]
		]
			
		\addplot table[
		x=Time,
		y=CLIENT_RESPONSE_TIME_Avg
		]{reference.dat};
		
%		\addplot[error bars/.cd,		
%		y dir=minus,
%		error mark=none,
%		error bar style={dotted}
%		] 
%		table[
%		x=Time,
%		y=CLIENT_RESPONSE_TIME_Avg,
%		y error=CLIENT_RESPONSE_TIME_Min,
%		]{reference.dat};

		\addplot table[
		x=Time, 
		y=OPTIMAL_RESPONSE_TIME_Avg
		]{reference.dat};

		\addplot table[
		x=Time, 
		y=REQUEST_PENDING_TIME_Avg
		]{reference.dat};

		\addplot table[
		x=Time, 
		y=CLIENT_RESPONSE_TIME_overtime
		]{reference.dat};

		\addplot table[
		x=Time, 
		y=OPTIMAL_RESPONSE_TIME_overtime
		]{reference.dat};

        \legend{{\footnotesize Response time}, {\footnotesize Opt. response time}, {\footnotesize Pending time}, {\footnotesize Avg response time}, {\footnotesize Avg opt. response time}}
        
        \end{axis}
    \end{tikzpicture}
       \caption{Average response time in reference experiment}
       \label{fig:reference_response_time}
\end{figure}
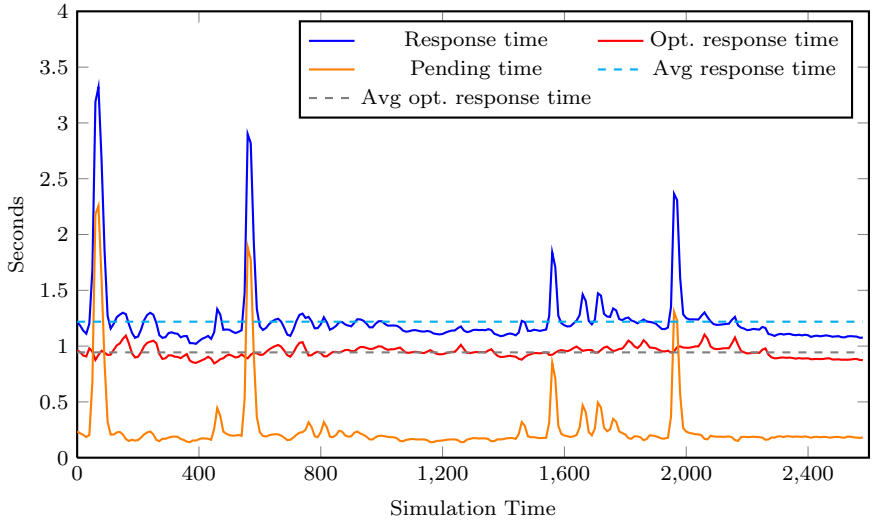

%\begin{figure}[htbp]
%    \begin{center}
%		\includegraphics[scale=0.48]{charts/reference_experiment/response_time} 
%        \caption{Average response time in reference experiment}
%        \label{fig:reference_response_time}
%    \end{center}
%\end{figure}

\begin{table}[htdp]
\begin{center}
\caption{Metrics for each scenario computed as average of time}
\label{tab:global_metrics}
\begin{tabular}{|p{2.3cm}|c|c|c|}
 \hline
 \textbf{Scenario} & \textbf{Average response} &  \textbf{Rejected} & \textbf{Lost} \\
 \textbf  & \textbf{time (sec)} &  \textbf{requests} & \textbf{requests} \\
 \hline
 Reference  & 1.23 & 0.36\% & - \\
 \hline
 Homogeneous & 1.24 & 0.34\% & - \\
 \hline
 Extremely unbalanced & 1.41 & 0.47\% & - \\
 \hline
 Churn soft & 1.48 & 0.72\% & 0.54\% \\
 \hline
 Churn heavy & 1.43 & 1.14\% & 1.10\% \\
 \hline
 Distruptive soft & 1.42 & 0.25\% & 0.08\% \\
 \hline
 Distruptive heavy & 1.46 & 0.5\% & 0.17\% \\
 \hline
 
\end{tabular}
\end{center}

\end{table}%

Starting from the reference experiment, we focused on quantifying the impact of different autonomic service capacities on the system behavior. In particular, we performed two experiments in which capacities were distributed according to two different configurations. 
Specifically, in the first experiment we set homogeneous capacities (i.e., $C_i = 1$ req/sec for every autonomic service), while in the second experiment capacities are set according to an extremely unbalanced configuration (i.e., 50\% with capacity $C_i = 0.1$ req/sec and 50\% with capacity $C_i = 1.9$ req/sec). 
As shown in Figure~\ref{fig:homogeneous_response_time}, average response time in the homogeneous setting is close to the one obtained in the reference scenario (see line 1 and 2 of Table \ref{tab:global_metrics}) despite the fact that capacities are different. The extremely unbalanced configuration, shown in Figure~\ref{fig:non_homogeneous_response_time}, reports an average response time of 1.41 sec, that is higher with respect to the previous case. This behavior is expected as requests have a higher probability to reach a service with a limited capacity and therefore experience longer waiting times before the actual processing. The 13\% increment of response time with respect to both reference and homogeneous scenario is however limited and shows that our approach is able to efficiently handle systems with highly different capacities.

%\documentclass[a4paper]{article}

%\usepackage{pgfplots}
%\pgfplotsset{compat=1.3}

%\begin{document}

\pgfplotsset{tick label style={
	font=\footnotesize},
    label style={font=\small},
    legend style={font=\small}
}
\pgfplotsset{compat=1.3}

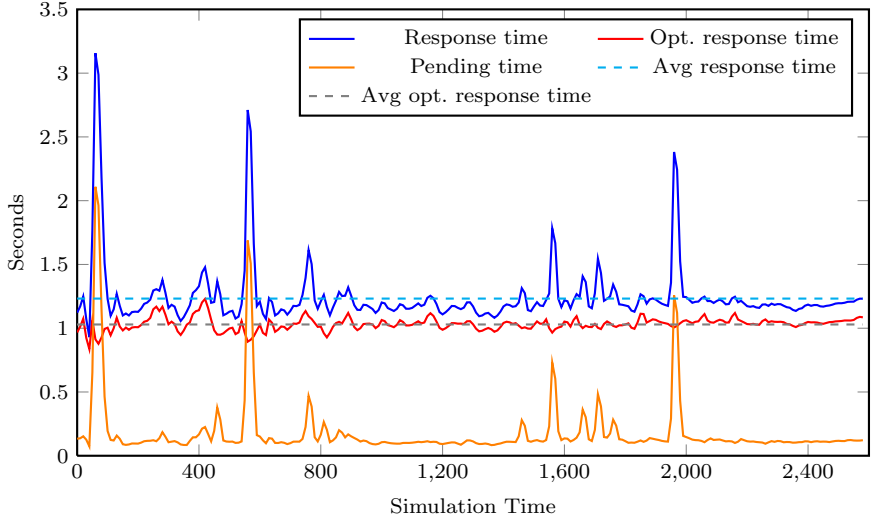
\begin{figure}
    \centering
    \begin{tikzpicture}
        \begin{axis}[
        small,
        samples=200,
		width=12cm,
		height=7.5cm,
		xlabel=Simulation Time,
		ylabel=Seconds,
		xmin=0,xmax=2600,ymin=0,ymax=3.5,
		xtick={0, 400,800,1200,1600,2000,2400,2800},
		cycle list={
			{blue},
			{red},
			{orange},
			{cyan,dashed},
			{gray,dashed}},
		thick,
		legend columns=2,
		legend style={/tikz/every even column/.append style={column sep=0cm, row sep=0cm}}]
		]
			
		\addplot table[
		x=Time,
		y=CLIENT_RESPONSE_TIME_Avg
		]{equal_capacity.dat};
							   
		\addplot table[
		x=Time, 
		y=OPTIMAL_RESPONSE_TIME_Avg
		]{equal_capacity.dat};

		\addplot table[
		x=Time, 
		y=REQUEST_PENDING_TIME_Avg
		]{equal_capacity.dat};

		\addplot table[
		x=Time, 
		y=CLIENT_RESPONSE_TIME_overtime
		]{equal_capacity.dat};

		\addplot table[
		x=Time, 
		y=OPTIMAL_RESPONSE_TIME_overtime
		]{equal_capacity.dat};

        \legend{{\footnotesize Response time}, {\footnotesize Opt. response time}, {\footnotesize Pending time}, {\footnotesize Avg response time}, {\footnotesize Avg opt. response time}}
        
        \end{axis}
    \end{tikzpicture}
        \caption{Average response time in the homogeneous capacities scenario}
        \label{fig:homogeneous_response_time}
\end{figure}
%\begin{figure}[htbp]
%    \begin{center}
			%\includegraphics[scale=0.48]{charts/homogeneous_capacities/response_time} 
%        \caption{Average response time in the homogeneous capacities scenario}
%        \label{fig:homogeneous_response_time}
%    \end{center}
%\end{figure}

%\documentclass[a4paper]{article}

%\usepackage{pgfplots}
%\pgfplotsset{compat=1.3}

%\begin{document}

\pgfplotsset{tick label style={
	font=\footnotesize},
    label style={font=\small},
    legend style={font=\small}
}
\pgfplotsset{compat=1.3}

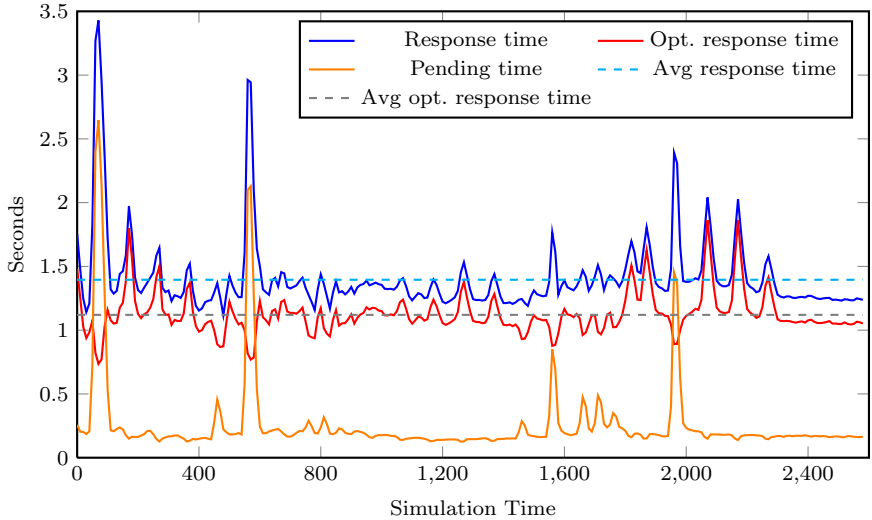
\begin{figure}
    \centering
    \begin{tikzpicture}
        \begin{axis}[
        small,
        samples=200,
		width=12cm,
		height=7.5cm,
		xlabel=Simulation Time,
		ylabel=Seconds,
		xmin=0,xmax=2600,ymin=0,ymax=3.5,
		xtick={0, 400,800,1200,1600,2000,2400,2800},
		cycle list={
			{blue},
			{red},
			{orange},
			{cyan,dashed},
			{gray,dashed}},
		thick,
		legend columns=2,
		legend style={/tikz/every even column/.append style={column sep=0cm, row sep=0cm}}]
		]
			
		\addplot table[
		x=Time,
		y=CLIENT_RESPONSE_TIME_Avg
		]{extreme_capacity.dat};
							   
		\addplot table[
		x=Time, 
		y=OPTIMAL_RESPONSE_TIME_Avg
		]{extreme_capacity.dat};

		\addplot table[
		x=Time, 
		y=REQUEST_PENDING_TIME_Avg
		]{extreme_capacity.dat};

		\addplot table[
		x=Time, 
		y=CLIENT_RESPONSE_TIME_overtime
		]{extreme_capacity.dat};

		\addplot table[
		x=Time, 
		y=OPTIMAL_RESPONSE_TIME_overtime
		]{extreme_capacity.dat};

        \legend{{\footnotesize Response time}, {\footnotesize Opt. response time}, {\footnotesize Pending time}, {\footnotesize Avg response time}, {\footnotesize Avg opt. response time}}
        
        \end{axis}
    \end{tikzpicture}
        \caption{Average response time in the extremely unbalanced capacities scenario}
        \label{fig:non_homogeneous_response_time}
\end{figure}

%\begin{figure}[htbp]
%    \begin{center}
%			\includegraphics[scale=0.48]{charts/non_homogeneous_capacities/response_time} 
%        \caption{Average response time in the extremely unbalanced capacities scenario}
%        \label{fig:non_homogeneous_response_time}
%    \end{center}
%\end{figure}

Another category of experiments we performed involves the introduction of external factors in the system such as the presence of failures. Indeed, real systems can be affected by unpredictable events compromising the functionality offered by the autonomic service (i.e., a software bug, malfunctioning hardware, etc.). We envision two possible scenarios: \emph{churn} and \emph{catastrophic}. 

In the churn scenario some autonomic services can fail at each time instant, thus creating the need for a continuous auto-scaling. When an autonomic service fails it loses all neighbor connections as well as existing requests in the queue and it is not able to process requests. 
We simulated churn according to two parameters: $p_{fail}$ is the probability of an autonomic service to fail and $T_{fail}$, the time period in which the probability is evaluated (for example $p_{fail} = 1\%$ and $T_{fail}=2$ means that each autonomic service can fail every two seconds with a probability of 1\%). 
We compare the behavior of the system in two possible configurations: \emph{soft churn} (with $p_{fail} = 5\%$ and  $T_{fail}=10$) and \emph{heavy churn} (with $p_{fail} = 10\%$ and  $T_{fail}=10$). 

\begin{figure}[htbp]
\centering
\subfigure[Response time with soft churn]{
\includegraphics[scale=0.48]{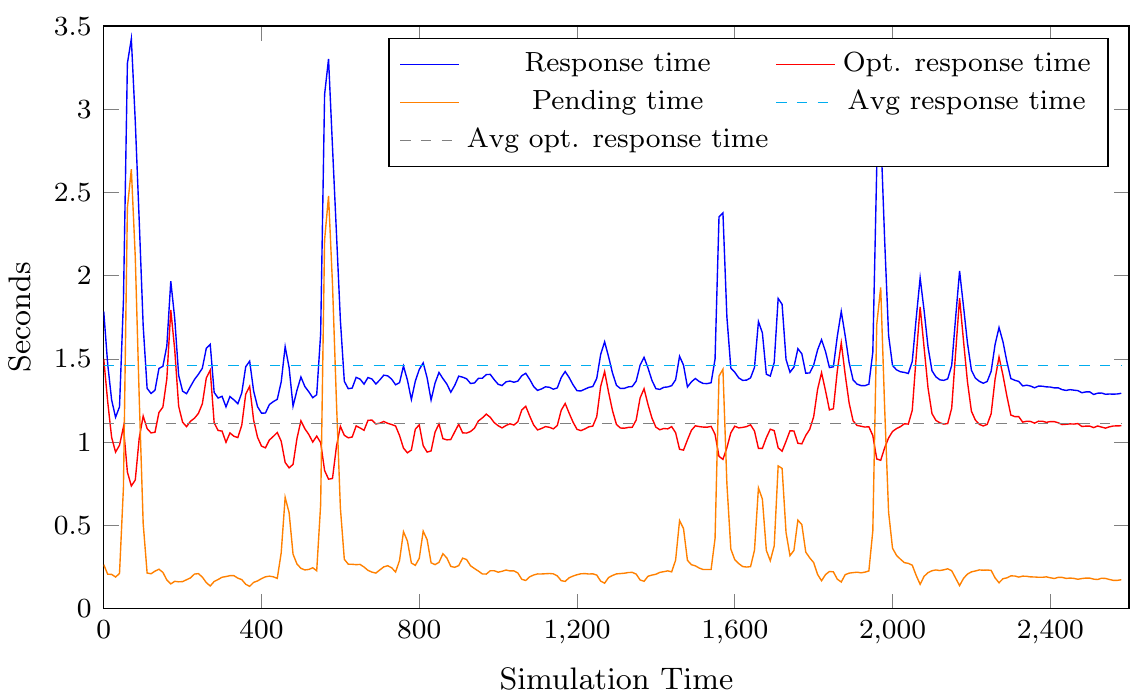}
\label{fig:churn_soft_response_time}
}
\subfigure[Response time with heavy churn]{
\includegraphics[scale=0.48]{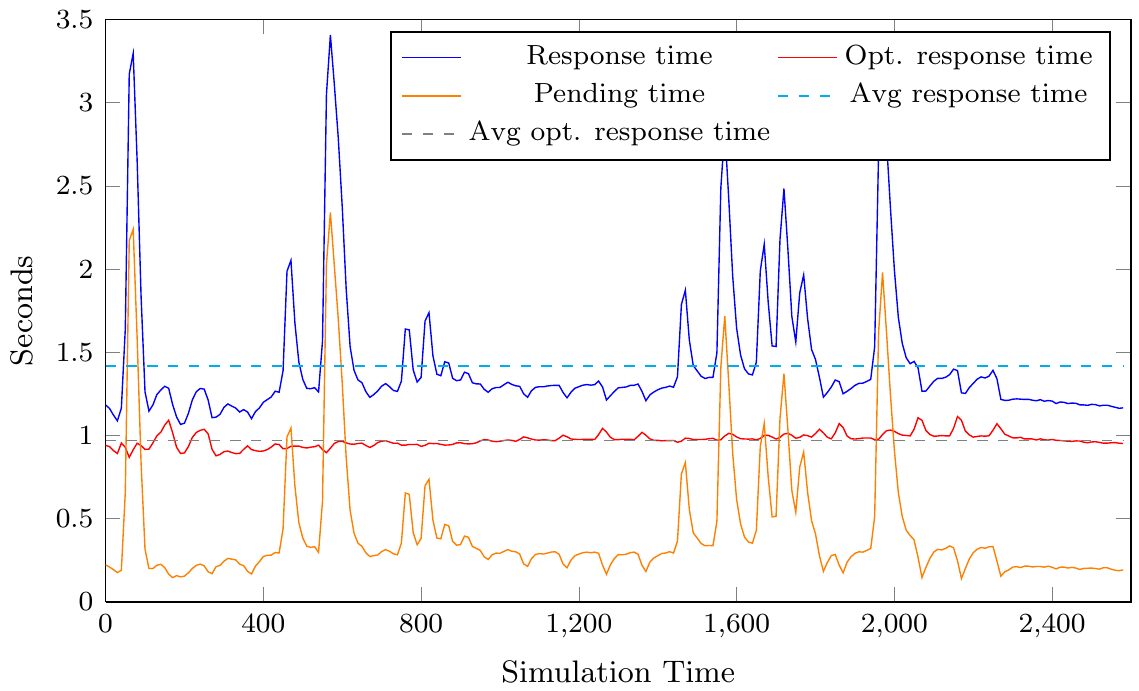}
\label{fig:churn_heavy_response_time}
}

\subfigure[Autonomic services with soft churn]{
\includegraphics[scale=0.47]{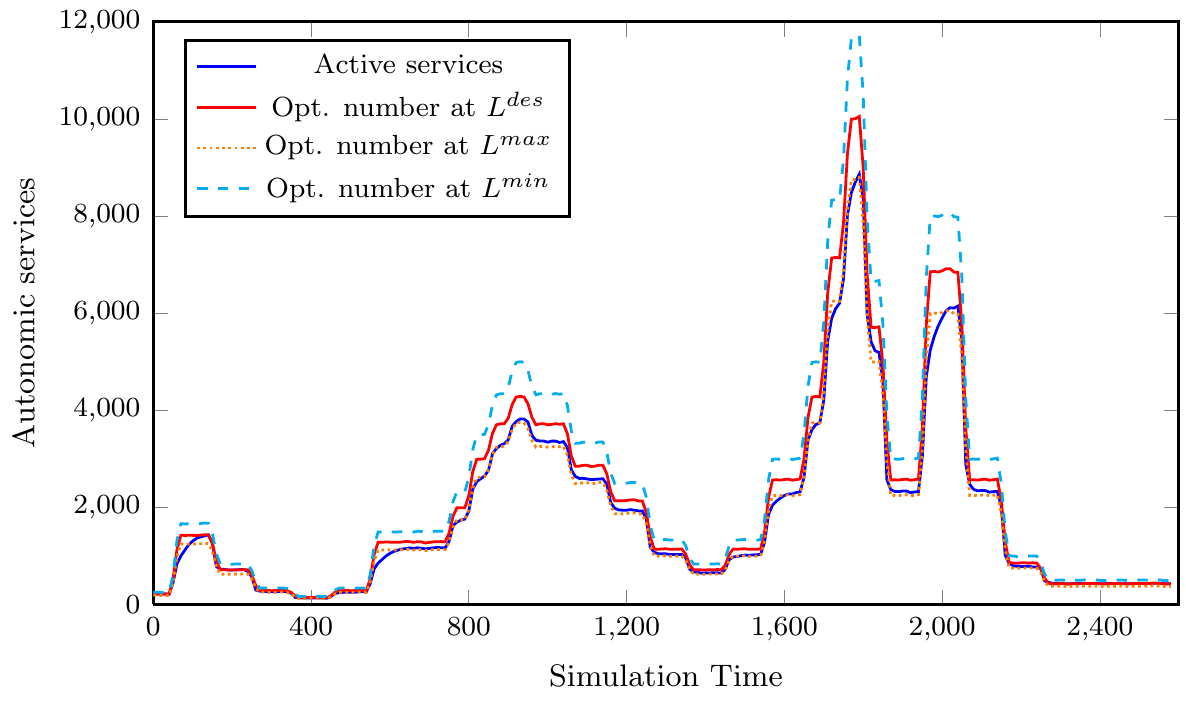}
\label{fig:churn_soft_number_of_nodes}
}
\subfigure[Autonomic services with heavy churn]{
\includegraphics[scale=0.47]{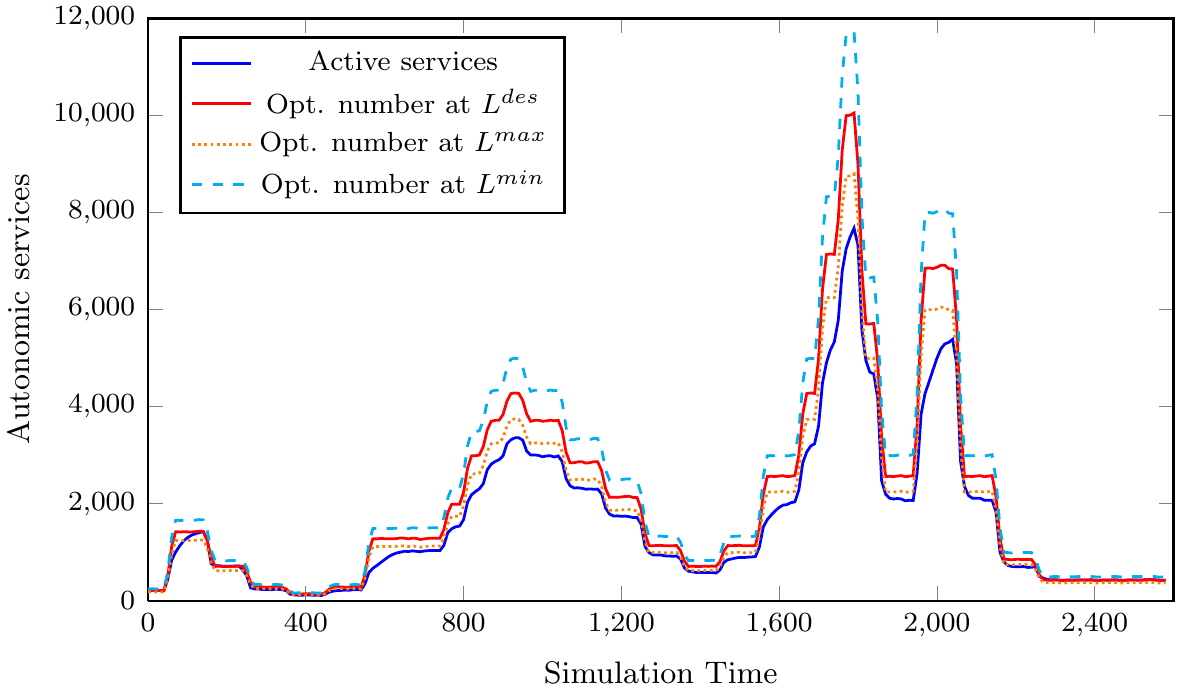}
\label{fig:churn_heavy_number_of_nodes}
}

\caption[Churn scenario]{Response time and number of autonomic services instantiated in soft ($p_{fail} = 5\%$, $T_{fail}=10$) and heavy ($p_{fail} = 10\%$, $T_{fail}=10$) \emph{churn} scenarios}
\label{fig:churn}

\end{figure}

As depicted in Figures \ref{fig:churn} \subref{fig:churn_soft_response_time} and \subref{fig:churn_heavy_response_time} the average response time is higher with respect to the reference scenario due to the presence of churn. In particular, as shown in line 4 and 5 of Table \ref{tab:global_metrics}, churn induces a higher amount of requests to be rejected or lost.
However, in the heavy churn scenario the response time is slightly lower than the one in the soft scenario. 
The apparent improvement is due to the fact that more requests are not processed (rejected or lost). As shown in Figure \ref{fig:churn} \subref{fig:churn_soft_number_of_nodes} and \subref{fig:churn_heavy_number_of_nodes}, the number of deployed services reflects this behavior. Specifically, the presence of churn impacts the precision of the system during high load peaks (i.e., sec 1700). Despite the stress imposed by heavy churn, our approach is still able to limit the number of rejected requests to 1.14\% of the total submitted requests.

Differently from the continuous churn, in the catastrophic scenario we tested our solution against a sudden critical event which may disrupt a substantial part of the data center infrastructure such as the one that happened in the Amazon infrastructure in April 2011 \cite{amazon-outage}. 

As for the case of churn, we structure the experiments in two configurations in which we vary the percentage of autonomic services killed: in \emph{soft} configuration 30\% of existing autonomic services are killed, while in \emph{heavy} configuration 60\% of the autonomic services are killed. 
Autonomic services to be killed are randomly chosen from the set of currently deployed autonomic services. We fixed the disruptive event to happen after the system has stabilized (sec 200) such that we can observe the effects of the disruption in the successive time instants. 
Simulations are conducted till sec 450, when all disruption effects are recovered by our algorithm and the system is stable.
As our interest is to understand how the system is able to recover after the disruptive event, we adopted a constant workload of 7200 req/sec. Adopting a dynamic workload as the one presented previously would make the quantification of the disruption less clear since the number of deployed services would change due to two external factors (workload and disruptive event).

%In order to isolate the possible effects on autonomic services of the input workload we have run the experiments with a constant workload of 7200 req/sec. 

\begin{figure}[htbp]
\centering
\subfigure[Number of autonomic services in the soft disruptive case]{
\includegraphics[scale=0.47]{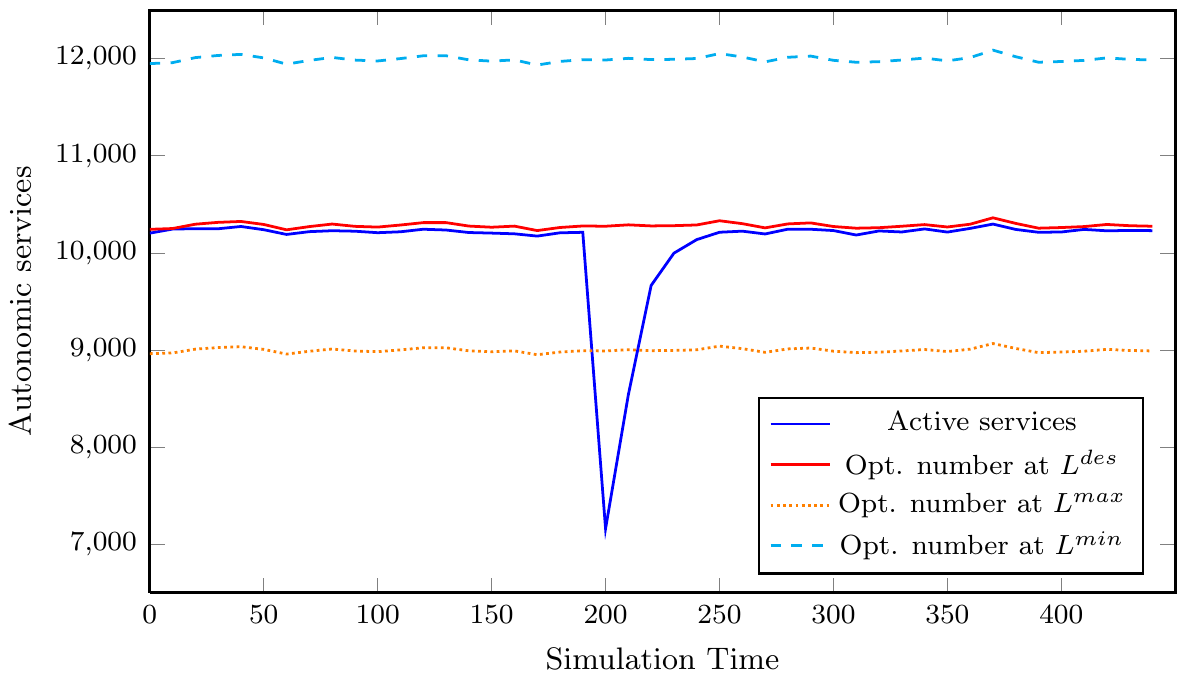}
\label{fig:disr_soft_number_of_nodes}
}
\subfigure[Number of autonomic services in the heavy disruptive case]{
\includegraphics[scale=0.47]{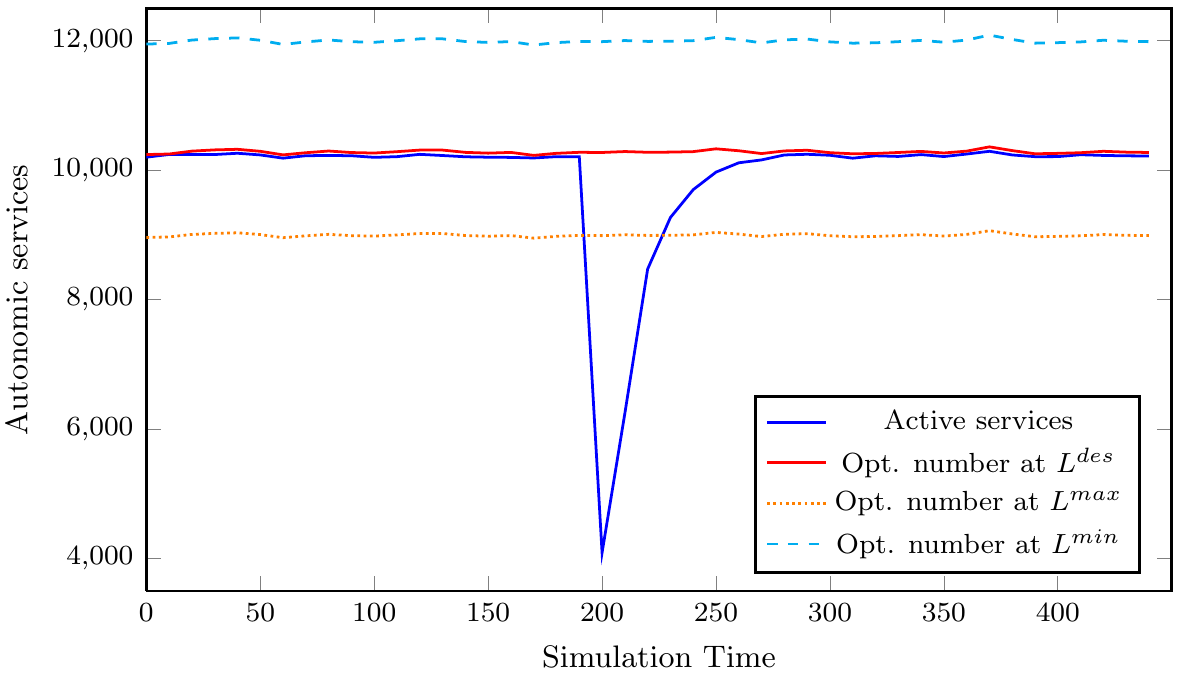}
\label{fig:disr_heavy_number_of_nodes}
}

\subfigure[Response time in the soft disruptive case]{
\includegraphics[scale=0.48]{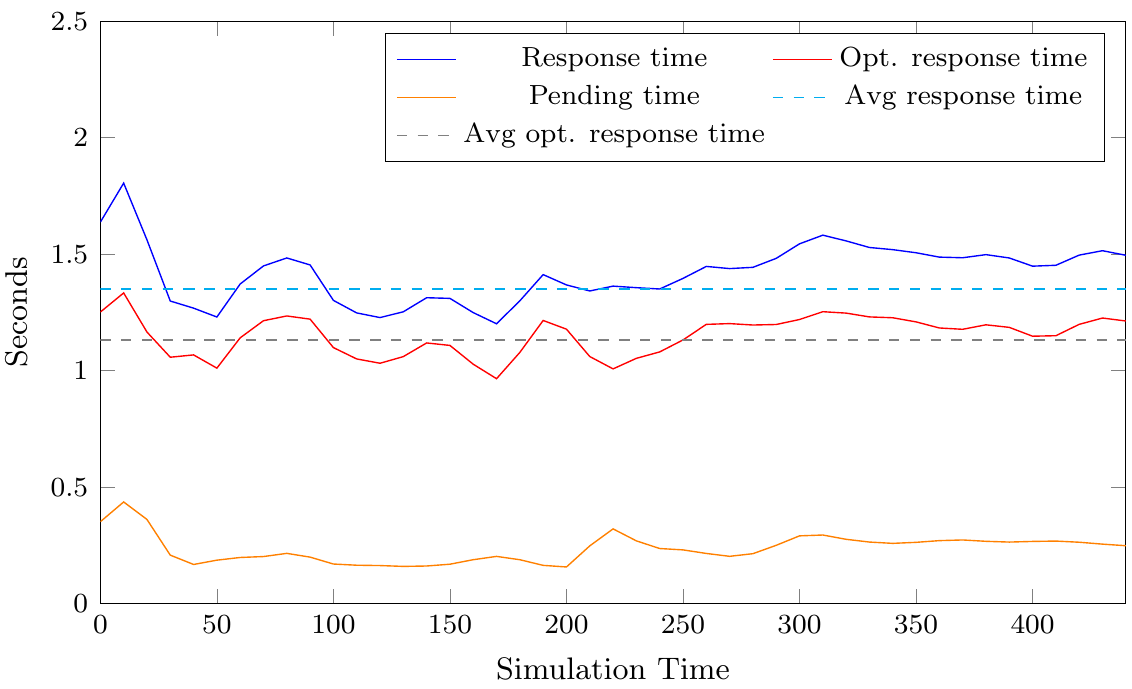}
\label{fig:disr_soft_response_time}
}
\subfigure[Response time in the heavy disruptive case]{
\includegraphics[scale=0.48]{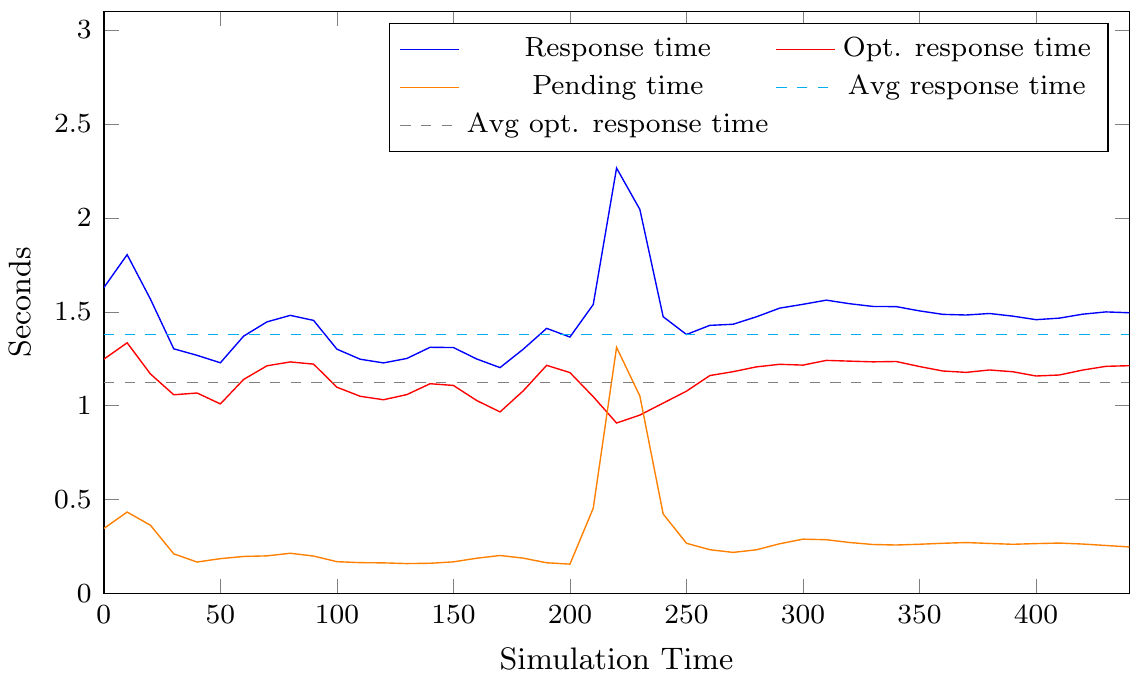}
\label{fig:disr_heavy_response_time}
}

\subfigure[Dropped and rejected requests in the soft disruptive case]{
\includegraphics[scale=0.48]{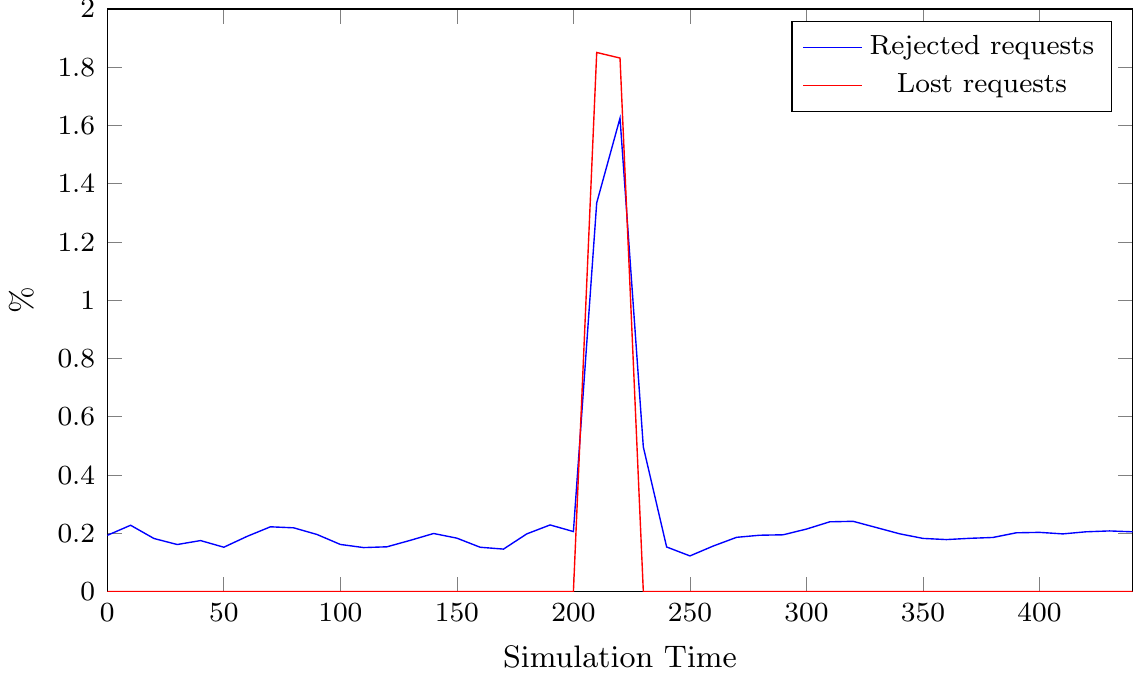}
\label{fig:disr_soft_drop_out}
}
\subfigure[Dropped and rejected requests in the heavy disruptive case]{
\includegraphics[scale=0.48]{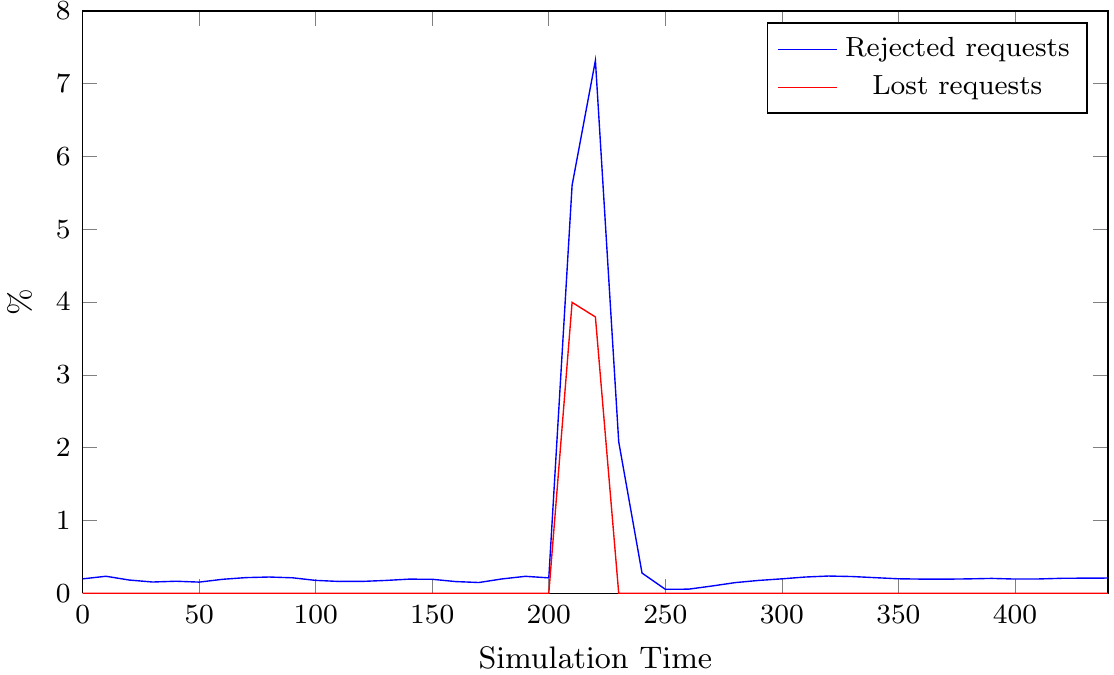}
\label{fig:disr_heay_drop_out}
}

\caption[Disruptive scenario]{Response time, number of autonomic services instantiated and dropped/rejected requests in soft and heavy \emph{disruptive} scenarios}
\label{fig:disruptive}

\end{figure}

The effects of the disruption in the deployed services are visible in Figures \ref{fig:disruptive} \subref{fig:disr_soft_number_of_nodes} and \subref{fig:disr_heavy_number_of_nodes}. In this case the system is able to recover from the damaged status in less than 50 seconds, by instantiating missing services till the previous amount of services is reached. The effects of failures at second 200 are visible in response time plots (see Figures~\ref{fig:disruptive} \subref{fig:disr_soft_response_time} and \subref{fig:disr_heavy_response_time}), where a small increment can be observed also for the pending time (services cannot efficiently accommodate all the incoming traffic). Finally, Figures~\ref{fig:disruptive} \subref{fig:disr_soft_drop_out} and \subref{fig:disr_heay_drop_out} show the amount of requests that is rejected and lost due to the disruptive event. In particular, up to 10\% of requests experience a rejection and 5\% experience a loss (these would be however lost using any other approach). Also for rejected requests the recovery is fast and in less than 50 seconds previous level of performance is restored.

% reference,churn heavy capacityClasses = 0.5:0.5,0.3:1.83,0.2:1
% no queue balancing, no admission,extreme capacity, distruptive soft, distruptive heavy, churn soft,  capacityClasses = 0.5:0.1,0.5:1.9
% equal capacity capacityClasses =1:1
% the number of initially active peers is always 200 except for experiments with distruptive churn (they also have a constant request rate).

\subsection{Comparisons and discussion}
\label{sub:discussion}

Here we discuss the achieved results and we carry out a comparison with the results obtained by other works in the same research area.
One relevant factor in evaluating the quality of a scaling algorithm is the scalability of the technique itself. By scalability we mean the ability of the technique to maintain its characteristics (i.e., response time, optimality of number of autonomic services, etc.) with increasing sizes of the problem (in terms of received requests per time unit and number of autonomic services needed). In this regard, the adoption of a highly variating workload trace ensures the validation of the technique at different sizes.

As previous charts show, the average response time experienced by clients is close to the optimal one in all experiments despite adverse experimental conditions such as churn and unbalanced capacities. Moreover, the quality of response time is not sacrificing the economic side; indeed, the number of active autonomic services is not far from the optimal one and, for the majority of the time instants, their load is within the defined thresholds. 
%defined by the values computed using the high and low thresholds for load. 

Another interesting property of our approach is the good level of reactivity (i.e., the ability to quickly re-adapt the number of autonomic services with sudden changes in the workload). However the adaptation is not instantaneous, as also shown by some peaks in the response time, and in case of massive change some requests can be lost (see Figure \ref{fig:referece_requests_cumulative} around second 2000). A high level of reactivity could however lead to instability phenomena (i.e., given a constant input the system continuously oscillates) or over/under provisioning; in our case we see that these phenomena are quite limited even in the case of external dynamism. 

Table \ref{table:test_sizes} compares our technique with the most relevant works already presented in Section~\ref{sec:state-of-the-art} concerning the problem of scalability in the cloud intended as addition and removal of computing units. Our system is the only one that has been validated over a large scale system while others remain under the threshold of 50 VMs. Similar to other works we report results with respect to the response time metric (relevant for clients) and the number of autonomic services allocated (relevant to service provider). However, differently from other works we also consider the optimality of our metrics in order to precisely assess how the system is behaving with respect to an optimal centralized system under ideal conditions always taking the optimal choice. Given the decentralized nature of our system, this last aspect is critical for different reasons that are not present in centralized systems: (i) the lack of a global knowledge puts each element in a condition of limited knowledge, (ii) each component of the system takes actions independently of others, (iii) communication delays among autonomic services can be source of imprecisions.

%As shown in Table \ref{table:test_sizes}, most works reports small to medium size experiments or simulations, while the only solution that addresses large scale systems beside the one we ours, is the work of Wuhib et al. \cite{wuhib2011}. Thus, the fact that their solution is scalable with respect to the number of virtual machines and the number of sites, but not with respect to the number of instances of a site can be deduced from the dimension of the simulations where huge numbers of servers and services were considered as opposed to a maximum of 100 service instances. 

\begin{table}[htbp]
\begin{center}
    \caption{Experimental settings adopted in similar works in the literature}
    \label{table:test_sizes}
\begin{tabular}{ | p{2cm} | p{3.5cm} | p{1.5cm} | p{3cm} |}
	\hline
	\textbf{Contribution} & \textbf{Evaluation metrics} & \textbf{Validation} & \textbf{Tests size} \\
	\hline
	Bonvin et al. \cite{bonvin2011} & Used resources, response time, throughput, SLA violations, Scale operations  & Experiment & 5 services, 8 servers, scaling up to 19 cores and 4 servers  \\
	\hline
	Iqbal et al. \cite{iqbal2011} & Throughput, response time, CPU utilization, SLA violations & Experiment & 2 services, 7 servers,  scaling up to 8 VMs \\ 
	\hline
	Meng et al. \cite{meng2010} & Throughput, number of instances, Convergence time, & Experiment & 1 service, scaling up to 50 VMs \\
	\hline
	Sharma et al. \cite{sharma2011}  & Infrastructure cost, response time, latency to actuate new configuration & Experiment & 2 services, 6 VMs \\
	\hline
	Xiong et al. \cite{xiong2011} & Response time, CPU utilization  & Experiment & 3 servers, 3 VMs \\
	\hline
%	Almeida et al. \cite{almeida2010}  & & Simulation & 1 server, 12 services/VMs \\
%	\hline
%	Ghanbari et al. \cite{ghanbari2012} & & Simulation & 7 servers, 10 services \\
%	\hline
%	Fox et al. \cite{fox1997} & & Experiment & 15 servers, 2 services \\
%	\hline
%	Wuhib et al. \cite{wuhib2011} & & Simulation & 160000 servers, 384000 services, 100 service instances \\
%	\hline
	Our solution & Response time (including optimality), number of instances (including optimality), dropped/rejected requests & Simulation & 10000 VMs \\ 
	\hline
\end{tabular}
\end{center}
\end{table}

Moreover, all reported techniques do not consider potential adversarial environment conditions due to dynamism in the network (i.e., faults). We believe that an efficient technique must consider this aspect, as faults are inevitable and already have put in crisis some existing cloud providers, as stated in \cite{amazon-outage}. To the best of our knowledge, our auto-scaling approach is the first decentralized one considering also network dynamism due to faults. On the side of our weaknesses we have the fact that, while other approaches have been validated on the actual system, even though on a very small scale, we have adopted simulations as the strategy of evaluation. We did so in order to analyze in a controlled environment large scale systems, but in a future work we plan to have a full validation on top of a real deployed system.

The shades around the experimental plot lines show that the developed solution is stable with respect to random factors maintaining a limited level of variability over many runs of the same experiment. This observation is not trivial as the technique itself involves decisions made according to probabilities computed at runtime (i.e., creation of a new autonomic service).
We also observe that the system is able to deal with a relevant number of autonomic services (i.e., the peak in our experiments reaches 10.000 VMs) while maintaining unaltered the quality of achieved results (response time optimality and optimality of autonomic services allocated), thus showing a good level of scalability for the technique itself.

\section{Conclusion}
\label{sec:conclusion}

In this paper we have presented \DEPAS{}, a decentralized probabilistic algorithm for scaling services in a cloud computing context. In particular, we propose a novel probabilistic auto-scaling algorithm, which, in combination with a robust overlay network among services and a decentralized load balancing technique, provides an effective decentralized solution for deploying massively scalable services in the cloud. The solution we have proposed will be suitable in all the situations in which a centralized solution is not feasible, like, for example, when dealing with multiple cloud providers in a federated cloud scenario.

We implemented our technique using Protopeer toolkit and validated it with an extensive set of simulations. Experiments show that \DEPAS{} scales to a large number of nodes (10.000) while maintaining a nearly-optimal response time and allocated resources; we stressed the system with a highly-variable workload trace inspired by a real scenario. Moreover, experiments have been conducted also taking into account potential dynamism in the system like continuous churn and disruptive events.

We are currently considering the extension of our system in two orthogonal directions: (i) support to service composition and (ii) multi-cloud optimizations. In the former direction the objective is to provide a scalability feature to applications composed of different services/components (for example a three-tier web system composed of a web-server, an application-server and a database-server); in this case the replication of the whole system might generate a waste of resources and the best decision would be to only consider a single service/component replication (for example only the web server). 
The second possible extension focuses on a multi-cloud context in which the auto-scaling needs to take into account other characteristics that are not currently captured by our solution, since they depend on the commercial strategy of each cloud. Examples of additional characteristics may include inter-cloud communication costs, and efficient heuristics to choose the best hosting cloud and the best type of instance (in terms of capacity) when replicating an autonomic service.

\begin{acknowledgements}
This research has been partially funded by the European Commission, under projects SMSCom (IDEAS-ERC 227977) and mOSAIC ( FP7-ICT-2009-5-256910) and by the Romanian National Authority for Scientific Research, CNCS Ð UEFISCDI, under project PN-II-ID-PCE-2011-3-0260 (AMICAS). The experimental part has been supported by Amazon AWS in Education research grant. Bogdan Caprarescu is partially supported by IBM through a PhD Fellowship Award. 
\end{acknowledgements}

\bibliographystyle{abbrv}
\bibliography{0-computing}

\begin{thebibliography}{10}

\bibitem{amazon_ec2}
Amazon elastic compute cloud (amazon ec2).
\newblock http://aws.amazon.com/ec2/ (accessed Sep 5 2011).

\bibitem{ankoder}
Ankoder.
\newblock http://www.ankoder.com/ (accessed Sep 5 2011).

\bibitem{town_hall}
Google developer products help whitehouse.gov connect with america.
\newblock
  http://googlecode.blogspot.com/2009/04/google-developer-products-help.html
  (accessed Sep 5 2011).

\bibitem{nimbusproject}
Nimbus project.
\newblock http://www.nimbusproject.org/ (accessed Sep 5 2011).

\bibitem{rightscale}
Rightscale.
\newblock http://www.rightscale.com/ (accessed Sep 5 2011).

\bibitem{scalarium}
Scalarium.
\newblock http://www.scalarium.com/ (accessed Sep 5 2011).

\bibitem{zencoder}
Zencoder.
\newblock http://zencoder.com/ (accessed Sep 5 2011).

\bibitem{auto-scaling}
Amazon auto scaling, 2011.
\newblock http://aws.amazon.com/autoscaling/ (accessed Sep 5 2011).

\bibitem{ec2}
Amazon ec2, 2011.
\newblock http://aws.amazon.com/ec2/ (accessed Sep 5 2011).

\bibitem{drools}
Drools, 2011.
\newblock http://www.jboss.org/drools/ (accessed Sep 5 2011).

\bibitem{amazon-outage}
Major amazon outage ripples across web, 2011.
\newblock
  http://www.datacenterknowledge.com/archives/2011/04/21/major-amazon-outage-ripples-across-web/
  (accessed Aug 7 2011).

\bibitem{adam2006}
C.~Adam and R.~Stadler.
\newblock A middleware design for large-scale clusters offering multiple
  services.
\newblock {\em IEEE Transactions on Network and Service Management},
  3(1):1--12, 2006.

\bibitem{almeida2010}
J.~Almeida, V.~Almeida, D.~Ardagna, I.~Cunha, C.~Francalanci, and M.~Trubian.
\newblock Joint admission control and resource allocation in virtualized
  servers.
\newblock {\em J. Parallel Distrib. Comput.}, 70:344--362, April 2010.

\bibitem{armbrust2010}
M.~Armbrust, A.~Fox, R.~Griffith, A.~D. Joseph, R.~Katz, A.~Konwinski, G.~Lee,
  D.~Patterson, A.~Rabkin, I.~Stoica, and M.~Zaharia.
\newblock A view of cloud computing.
\newblock {\em Commun. ACM}, 53:50--58, April 2010.

\bibitem{baresi2011}
L.~Baresi and S.~Guinea.
\newblock A3: self-adaptation capabilities through groups and coordination.
\newblock In {\em Proceedings of the 4th India Software Engineering
  Conference}, ISEC '11, pages 11--20, New York, NY, USA, 2011. ACM.

\bibitem{bolch}
G.~Bolch, S.~Greiner, H.~de~Meer, and K.~S. Trivedi.
\newblock {\em Queueing networks and Markov chains: modeling and performance
  evaluation with computer science applications}.
\newblock Wiley-Interscience, New York, NY, USA, 1998.

\bibitem{bonvin2011}
N.~Bonvin, T.~G. Papaioannou, and K.~Aberer.
\newblock Autonomic sla-driven provisioning for cloud applications.
\newblock In {\em Proceedings of the 2011 11th IEEE/ACM International Symposium
  on Cluster, Cloud and Grid Computing}, pages 434--443, 2011.

\bibitem{caprarescu2009}
B.~A. Caprarescu and D.~Petcu.
\newblock A self-organizing feedback loop for autonomic computing.
\newblock In {\em Proceedings of the 2009 Computation World: Future Computing,
  Service Computation, Cognitive, Adaptive, Content, Patterns},
  COMPUTATIONWORLD '09, pages 126--131, Washington, DC, USA, 2009. IEEE
  Computer Society.

\bibitem{celesti2010}
A.~Celesti, F.~Tusa, M.~Villari, and A.~Puliafito.
\newblock How to enhance cloud architectures to enable cross-federation.
\newblock In {\em Proceedings of the 2010 IEEE 3rd International Conference on
  Cloud Computing}, CLOUD '10, pages 337--345, Washington, DC, USA, 2010. IEEE
  Computer Society.

\bibitem{dinitto2008}
E.~Di~Nitto, D.~J. Dubois, R.~Mirandola, F.~Saffre, and R.~Tateson.
\newblock Applying self-aggregation to load balancing: experimental results.
\newblock In {\em Proceedings of the 3rd International Conference on
  Bio-Inspired Models of Network, Information and Computing Sytems}, BIONETICS
  '08, pages 14:1--14:8. ICST (Institute for Computer Sciences,
  Social-Informatics and Telecommunications Engineering), 2008.

\bibitem{fox1997}
A.~Fox, S.~D. Gribble, Y.~Chawathe, E.~A. Brewer, and P.~Gauthier.
\newblock Cluster-based scalable network services.
\newblock In {\em Proceedings of the sixteenth ACM symposium on Operating
  systems principles}, SOSP '97, pages 78--91, New York, NY, US, 1997. ACM.

\bibitem{frincu2011}
M.~Fr\^{\i}ncu, N.~M. Villegas, D.~Petcu, H.~A. M{\"u}ller, and R.~Rouvoy.
\newblock Self-healing distributed scheduling platform.
\newblock In {\em CCGRID}, pages 225--234, 2011.

\bibitem{buyya2010}
M.~E. Frincu, N.~M. Villegas, D.~Petcu, H.~A. Muller, and R.~Rouvoy.
\newblock Self-healing distributed scheduling platform.
\newblock In {\em Proceedings of the 2011 11th IEEE/ACM International Symposium
  on Cluster, Cloud and Grid Computing}, CCGRID '11, pages 225--234,
  Washington, DC, USA, 2011. IEEE Computer Society.

\bibitem{protopeer}
W.~Galuba, K.~Aberer, Z.~Despotovic, and W.~Kellerer.
\newblock Protopeer: a p2p toolkit bridging the gap between simulation and live
  deployement.
\newblock In {\em Proceedings of the 2nd International Conference on Simulation
  Tools and Techniques}, Simutools '09, pages 60:1--60:9, ICST, Brussels,
  Belgium, Belgium, 2009.

\bibitem{garlan2004}
D.~Garlan, S.-W. Cheng, A.-C. Huang, B.~Schmerl, and P.~Steenkiste.
\newblock Rainbow: Architecture-based self-adaptation with reusable
  infrastructure.
\newblock {\em Computer}, 37:46--54, October 2004.

\bibitem{ghanbari2012}
H.~Ghanbari, B.~Simmons, M.~Litoiu, and G.~Iszlai.
\newblock Feedback-based optimization of a private cloud.
\newblock {\em Future Generation Comp. Syst.}, 28(1):104--111, 2012.

\bibitem{iqbal2011}
W.~Iqbal, M.~N. Dailey, D.~Carrera, and P.~Janecek.
\newblock Adaptive resource provisioning for read intensive multi-tier
  applications in the cloud.
\newblock {\em Future Generation Computer Systems}, 27(6):871 -- 879, 2011.

\bibitem{Jelasity2007}
M.~Jelasity, S.~Voulgaris, R.~Guerraoui, A.-M. Kermarrec, and M.~van Steen.
\newblock Gossip-based peer sampling.
\newblock {\em ACM Trans. Comput. Syst.}, 25, August 2007.

\bibitem{keahey2009sky}
K.~Keahey, M.~Tsugawa, A.~Matsunaga, and J.~Fortes.
\newblock Sky computing.
\newblock {\em Internet Computing, IEEE}, 13(5):43--51, 2009.

\bibitem{kephart2003}
J.~O. Kephart and D.~M. Chess.
\newblock The vision of autonomic computing.
\newblock {\em Computer}, 36:41--50, January 2003.

\bibitem{kephart20}
J.~O. Kephart and W.~E. Walsh.
\newblock An artificial intelligence perspective on autonomic computing
  policies.
\newblock In {\em POLICY}, pages 3--12, 2004.

\bibitem{leadsopennebula}
O.~Leads.
\newblock Opennebula: The open source toolkit for cloud computing.
\newblock http://opennebula.org/ (accessed Sep 5 2011).

\bibitem{meng2010}
S.~Meng, L.~Liu, and V.~Soundararajan.
\newblock Tide: achieving self-scaling in virtualized datacenter management
  middleware.
\newblock In {\em Proceedings of the 11th International Middleware Conference
  Industrial track}, Middleware Industrial Track '10, pages 17--22, New York,
  NY, USA, 2010. ACM.

\bibitem{milojicic2002peer}
D.~Milojicic, V.~Kalogeraki, R.~Lukose, K.~Nagaraja, J.~Pruyne, B.~Richard,
  S.~Rollins, and Z.~Xu.
\newblock Peer-to-peer computing.
\newblock 2002.

\bibitem{nurmi2009eucalyptus}
D.~Nurmi, R.~Wolski, C.~Grzegorczyk, G.~Obertelli, S.~Soman, L.~Youseff, and
  D.~Zagorodnov.
\newblock The eucalyptus open-source cloud-computing system.
\newblock In {\em Proceedings of the 2009 9th IEEE/ACM International Symposium
  on Cluster Computing and the Grid}, pages 124--131. IEEE Computer Society,
  2009.

\bibitem{petcu2011}
D.~Petcu, C.~Cr\v{a}ciun, M.~Neagul, M.~Rak, and I.~Lazcanotegui.
\newblock Building an interoperability {API} for {Sky Computing}.
\newblock In {\em Proceedings of 2011 International Conference on High
  Performance Computing and Simulation (HPCS)}, pages 405--411. IEEE Computer
  Press, 2011.

\bibitem{sharma2011}
U.~Sharma, P.~Shenoy, S.~Sahu, and A.~Shaikh.
\newblock A cost-aware elasticity provisioning system for the cloud.
\newblock In {\em Proceedings of the 2011 31st International Conference on
  Distributed Computing Systems}, ICDCS '11, pages 559--570, Washington, DC,
  USA, 2011. IEEE Computer Society.

\bibitem{venticinque2010}
S.~Venticinque, R.~Aversa, B.~Di~Martino, M.~Rak, and D.~Petcu.
\newblock A cloud agency for sla negotiation and management.
\newblock In {\em Proceedings of the 2010 conference on Parallel processing},
  Euro-Par 2010, pages 587--594, Berlin, Heidelberg, 2011. Springer-Verlag.

\bibitem{weyns2008}
D.~Weyns, R.~Haesevoets, B.~Van~Eylen, A.~Helleboogh, T.~Holvoet, and
  W.~Joosen.
\newblock Endogenous versus exogenous self-management.
\newblock In {\em Proceedings of the 2008 international workshop on Software
  engineering for adaptive and self-managing systems}, SEAMS '08, pages 41--48,
  New York, NY, USA, 2008. ACM.

\bibitem{wuhib2010}
F.~Wuhib, R.~Stadler, and M.~Spreitzer.
\newblock Gossip-based resource management for cloud environments.
\newblock In {\em Proceedings of the 2010 International Conference on Network
  and Service Management (CNSM)}, pages 1--8, 2010.

\bibitem{xiong2011}
P.~Xiong, Z.~Wang, S.~Malkowski, Q.~Wang, D.~Jayasinghe, and C.~Pu.
\newblock Economical and robust provisioning of n-tier cloud workloads: A
  multi-level control approach.
\newblock In {\em Proceedings of the 2011 31st International Conference on
  Distributed Computing Systems}, ICDCS '11, pages 571--580, Washington, DC,
  USA, 2011. IEEE Computer Society.

\end{thebibliography}

\end{document}